\documentclass[fleqn,usenatbib]{mnras}
\usepackage{newtxtext,newtxmath}
\usepackage[T1]{fontenc}
\DeclareRobustCommand{\VAN}[3]{#2}
\let\VANthebibliography\thebibliography
\def\thebibliography{\DeclareRobustCommand{\VAN}[3]{##3}\VANthebibliography}
\usepackage{graphicx}
\usepackage{amsmath}
\usepackage{color}
\usepackage[dvipsnames]{xcolor}
\usepackage{subcaption}

\newcommand{\M}{\mbox{$M_{\odot}$}}
\newcommand{\rs}{\,\mathrm{rad}\,\mathrm{s}^{-1}}
\newcommand{\cm}{\mbox{cm}}
\newcommand{\g}{\mbox{g}}
\newcommand{\s}{\mbox{s}}

\title[\texttt{GRaM-X} GWs]{Gravitational waves from magnetorotational core-collapse supernovae using 3D GRMHD simulations: effect of rotation and magnetic fields}

\author[S. C. Schnauck]{Sophia C.~Schnauck, $^{1,2}$\thanks{E-mail: \href{mailto:schnauck@mpa-garching.mpg.de}{schnauck@mpa-garching.mpg.de}}
Swapnil Shankar,$^{3,4,2}$
Philipp M\"osta,$^{5,2,6}$
Roland Haas,$^{7, 8, 9}$
Erik Schnetter$^{10, 11, 12}$
\\
$^{1}$ Max-Planck-Institut f\"ur Astrophysik, Karl-Schwarzschild-Str. 1, D-85748 Garching, Germany\\
$^{2}$ API, University of Amsterdam, Science Park 904, 1098 XH Amsterdam, Netherlands\\
$^{3}$ Faculty of Mathematics, Informatics and Natural Sciences, University of Hamburg, Gojenbergsweg 112, 21029 Hamburg, Germany \\
$^{4}$Department of Physics \& Astronomy, University of Tennessee, Knoxville, Tennessee 37996, United States\\
$^{5}$GRAPPA, University of Amsterdam, Science Park 904, 1098 XH Amsterdam, Netherlands\\
$^{6}$IoP, University of Amsterdam, Science Park 904, 1098 XH Amsterdam, Netherlands \\
$^{7}$ Department of Phyiscs and Astronomy, University of British Columbia, 325 - 6224 Agricultural Road, Vancouver, BC, Canada \\
$^{8}$ National Center for Supercomputing applications, University of Illinois, 1205 W Clark St, Urbana, Illinois, USA \\
$^{9}$ Department of Physics, University of Illinois, 1110 West Green St, Urbana, Illinois, USA\\
$^{10}$ Perimeter Institute for Theoretical Physics, Waterloo, Ontario, Canada\\ 
$^{11}$ Department of Physics and Astronomy, University of Waterloo, Waterloo, Ontario, Canada\\ 
$^{12}$ Center for Computation \& Technology, Louisiana State University, Baton Rouge, Louisiana, USA
}

\date{Accepted XXX. Received YYY; in original form ZZZ}

\pubyear{\the\year{}}

\begin{document}
\label{firstpage}
\pagerange{\pageref{firstpage}--\pageref{lastpage}}
\maketitle

\begin{abstract}
We investigate the gravitational wave emission for 10 supernova progenitors from magnetorotational core-collapse to the supernova explosion using fully three-dimensional dynamical-spacetime general-relativistic magnetohydrodynamics simulations with the GPU-accelerated code \texttt{GRaM-X}. We consider 2 progenitors of zero-age-main-sequence mass $25\M$ and 8 with zero-age-main-sequence masses of $35\M$. For these models, we explore a range of rotation rates between $0.0$ and $3.5 \rs$, along with initial seed magnetic field of either $10^{12}\mathrm{G}$ or $10^{13}\mathrm{G}$. The analysis of the 10 models presented provides a comprehensive and systematic initial investigation of the interplay between progenitor rotation, magnetic field strength, and progenitor structure in shaping the explosion dynamics and gravitational wave (GW) emission. We find that stronger seed magnetic fields ($10^{13}\mathrm{G}$) suppress the GW strain amplitude relative to models with weaker initial fields ($10^{12}\mathrm{G}$). Increasing the initial rotation rate results in a more dynamical explosion, yielding correspondingly stronger gravitational waves. In addition, the progenitor mass/composition also exhibit a significant impact on the explosion dynamics and the morphology of the resulting waveforms. Finally, we find that all of our models lie above the detectability threshold for 3rd generation detectors aLIGO, Einstein Telescope, and Cosmic explorer at a $10\mathrm{kpc}$ distance and most would even still be detectable at $10\mathrm{Mpc}$, opening the possibility for observing gravitational wave emission for CCSNe beyond our galaxy. 
\end{abstract}

\begin{keywords}
supernovae -- gravitational waves -- magnetohydrodynamics -- stars: rotation 
\end{keywords}

\section{Introduction}
\label{sec:intro}

The gravitational collapse of massive stars ($M \gtrapprox 8\, \M$ at zero-age main-sequence (ZAMS)) may result in a supernova explosion, releasing a vast amount of gravitational binding energy ($\sim\mathcal{O}(10^{53})\,\mathrm{erg} = \mathcal{O}(1) \mathrm{B}$). The majority of this energy ($\sim 99\%$) is carried away by neutrinos, with only about $\sim 1\%$ driving the supernova explosion. Supernovae with energies $\mathcal{O}(1) B$ may be powered by the neutrino-driven mechanism \citep{BetheWilson1985}, wherein neutrinos emitted from the cooling proto-neutronstar (PNS) deposit energy into the gain region (the region behind the stalled shock wave), thus re-energizing it and enabling it to propagate outward \citep{Janka_Neutrino_Mechanism, NeutrinoDriven2_Müller, NeutrinoDriven_Vartanyan}.

While the neutrino-driven mechanism is thought to be responsible for a majority of CCSNe, especially those of non-rotating progenitors with weak magnetic fields, there exists a subclass - broad-lined type-Ic supernovae (SNe Ic-bl, also referred to as ``Hypernovae"). The energy of these Hypernovae is about 10 times greater than that of typical supernovae and cannot be explained by the neutrino-driven mechanism  \citep{Janka_2007, Burrows_2021}. For such extreme explosions, the magnetorotational mechanism has been proposed \citep{LeBlanc_Wilson_1970, Bisnovatyi_Kogan_1971, Meier1976, Wheeler_2002}. For this mechanism, rapid core rotation (with periods of $\mathcal{O}(1)\, \mathrm{ms}$) and magnetar-strength magnetic fields ($\sim 10^{15}\, \mathrm{G}$)  dominated by toroidal components are required, launching a bipolar, jetted explosion \citep{Burrows2007}. 

Although the vast majority of massive stars are expected to have slowly rotating cores, a small fraction, approximately $1\%$ \citep{WoosleyHeger2006},  may possess sufficient angular momentum to enable a magnetorotational explosion. These rapidly rotating progenitors have been proposed to originate from binary mergers where, for example, mass accretion can lead to angular momentum transfer between donor and accretor stars, leading to an increase in rotation \citep{Packet1981_RapidRotation}. Another proposed mechanism which may result in a rapidly rotating massive pre-supernova star is chemically homogeneous evolution \citep{Chemically_Homogeneous_Evolution_deMink_2010}. 

Several mechanisms for the emergence of a magnetic field in a supernova progenitor have been suggested. One possibility is binary interactions, particularly during a common envelope phase \citep{MagneticFieldsMassiveStars}. Another pathway by which the magnetic field can be amplified is turbulent shear, which leads to an increase in the magnetic pressure along the rotation axis. The combination of rapid rotation (winding the magnetic field lines) and convective turbulence can then drive dynamo processes \citep{Thompson_1993,Raynaud2020, Masada_2022} and together with the MRI further increase magnetic field strengths to values on the order of $10^{12}\mathrm{G}$ \citep{Balbus1991, Akiyama_2003, Reboul-Salze_2021, Obergaulinger_2009, MRI, Moesta_2015_Dynamo}.

Magnetorotational CCSNe are of particular interest as prominent sources of neutrinos and GWs as they can produce the strongest supernova GW emission, making them more easily detectable. The energy of GW emission from CCSNe is in the range of $\sim\mathcal{O}(10^{46}) - \mathcal{O}(10^{47})\mathrm{erg}$ within the first second after core-collapse \citep{Scheidegger_E_GW}, and thus two orders of magnitude lower than the total ejecta energy and the energy emitted across the electromagnetic spectrum \citep{SNReview}. As light cannot escape from the highly opaque stellar material until much after the explosion has occurred, GWs provide a unique way to gain insights into the multi-dimensional dynamics of the collapse and subsequent explosion of the star, the structural and compositional evolution of the PNS and the rotational profile of the collapsed core as they carry information from the innermost regions of the star. Most importantly, gravitational wave observations can provide critical insights into the explosion mechanism itself \citep{Ott_2009}. 

The GW emission from a CCSN has been shown to be highly dependent on the initial conditions of the progenitor, particularly the rotation rate and magnetic field configuration~\citep{Bugli2023, Kuroda_2021, Shibagaki_GWs_2024, OttAbdikamalov_2012}. Modeling GWs from rotating core-collapse and core-bounce requires a multi-physics approach: ideally the equations of general relativity and MHD, as well as a microphysical equation of state and neutrino transport need to be coupled in full 3D simulations. Recent developments in software development and supercomputing hardware have allowed us to employ a GPU accelerated code to tackle this complex problem and have made longer and more complex simulations more feasible. This work in particular has been enabled by our newly-developed GPU-accelerated dynamical-spacetime ideal-GRMHD code \texttt{GRaM-X} \citep{Gram-X}. 

In this study, we present gravitational wave results from 10 core-collapse supernova models with varying initial rotation rates between  $0.0$ and $3.5 \rs$, 2 different seed magnetic field strengths, and 2 progenitor models. We investigate the impact of these progenitor parameters on the resulting explosion dynamics and GW waveforms. To the best of our knowledge, this work constitutes the largest set of 3D GRMHD simulations of GWs from magnetorotationally-driven CCSNe.

The structure of the paper is as follows. We describe the numerical setup and methods in section~\ref{sec:numerics}. We give an overview of the dynamical evolution of the 10 progenitors in section~\ref{sec:pb_dyn}. In section~\ref{sec:GWs}, we show the evolution of the gravitational wave signatures. Finally, we discuss the results and present our conclusions in section~\ref{sec:conclusion}.

\section{Numerical Setup and Methods}
\label{sec:numerics}

We perform simulations in ideal GRMHD using the numerical relativity code infrastructure of the \texttt{Einstein Toolkit}. In particular, we employ the newly developed \texttt{GRaM-X} (\textbf{G}eneral \textbf{R}elativistic \textbf{a}ccelerated \textbf{M}agnetohydrodynamics on AMRe\textbf{X}) code~\citep{Gram-X}. We use the Z4c formulation to solve Einstein's equations and evolve the ideal GRMHD equations in a finite-volume fashion via the Valencia formulation. We use WENO5 reconstruction \citep{WENO5_1,WENO5_2} with fallback to 2nd-order TVD reconstruction \citep{Toro1999} in regions where WENO5 yields unphysical values (e.g. in regions of extreme dynamics), the HLLE Riemann solver \citep{HLLE}, and constrained transport \citep{Toth} to enforce the divergence-free condition ($\text{div}(\textbf{B}) = 0$) for the magnetic field. 

We adopt a finite-temperature microphysical equation of state (EOS) for the simulations, specifically the \textit{LS220} EOS from  \citep{LS220}, along with an approximate neutrino transport (M0 scheme) following~\citep{Radice_2016, Radice_2018}. The computational grid consists of a total of 9 AMR levels. Before collapse, we begin the simulations with 4 AMR levels with base grid boundaries at $\sim \pm 12000 \mathrm{km}$ in each direction and a coarse resolution of $94\mathrm{km}$. We increase the resolution by a factor of 2 for each new refinement level. As the core begins to collapse, we gradually add 5 more refinement levels when the central density exceeds $8\times10^{10}$, $3.2\times10^{11}$,$ 1.28\times10^{12}$, $5.12\times10^{12}$, and $2.048\times10^{13} \g\;\cm^{-3}$, with extents of $\sim \pm 190 \mathrm{km}$, $\sim \pm 120\mathrm{km}$, $\sim \pm 60\mathrm{km}$, $\sim \pm 30\mathrm{km}$, and $\sim \pm 20\mathrm{km}$, respectively. Once the shock begins expanding, the code dynamically increases the extent of refinement level 6 (which has resolution $\sim1.48\, \mathrm{km}$) so that the shock is always contained within this level and is thus resolved with $\sim1.48\, \mathrm{km}$.

We use 2 different pre-collapse progenitor models for the simulations, one with a zero-age main sequence (ZAMS) mass of $25\,\M$ and solar metallicity~\citep{E25}, and the other with a ZAMS mass of $35 \M$ and $10\%$ solar metalicity~\citep{35OC}. In total we also investigate two different magnetic field configurations and 6 rotation rates. The 10 models are labeled \texttt{xxM\_RyyBzz}, where \texttt{xx} represents the stellar ZAMS mass, \texttt{yy} the initial central angular velocity $\Omega_0 = yy/10 \rs$, and \texttt{zz} denotes the seed magnetic field strength of $10^{zz}\,\mathrm{G}$.

We impose an axisymmetric precollapse rotation profile on all models using the rotation law of~\citep{RotLaw_1991, Takiwaki_rotation_law}:
\begin{align}
    \Omega =  \Omega_0 \dfrac{r_0^2}{r_0^2 + r^2}\dfrac{z_0^4}{z_0^4 + z^4}, 
\end{align}
where $r_0 = 500\,\mathrm{km}$ implies moderate differential rotation and $z_0 = 1000\,\mathrm{km}$ implies low vertical shear. We will henceforth refer to $\Omega_0$ as the rotation rate of our model. Table \ref{table:model_details} shows the rotation rates for all models. Apart from the non-rotating model ($\Omega_0 = 0$), we choose rotation rates in the range $0.7\leq\Omega_0 \leq 3.5\, \mathrm{rad/s}$ which correspond to initial pre-collapse rotation periods of $8.98$ to $1.80\, \mathrm{s}$, respectively. 

As explained in Section~\ref{sec:intro}, an initially weak magnetic field may be amplified significantly by methods such as turbulent shears along with dynamo processes and the MRI. However, since resolving the MRI poses a significant computational challenge, our simulations are initiated with unrealistically strong initial magnetic field strengths under the assumption that the MRI will amplify the magnetic field to the dynamically important value of $\sim10^{15}-10^{16}\mathrm{G}$. We impose a pre-collapse magnetic field which is nearly uniform and parallel to the rotational axis inside the core and dipolar outside the core using a vector potential of the form
\begin{align}
    A_r &= A_\theta = 0; \\
    A_\phi &= B_0 (r_0^3)(r^3 + r_0^3)^{-1} r \sin\theta, 
\end{align}
where $B_0$ is a model constant for the initial seed field strength \citep{Takiwaki_bfield}. We initialize the progenitors with a field strength of either $10^{12}\mathrm{G}$ (henceforth referred to as $\texttt{B12}$) or $10^{13}\mathrm{G}$ (henceforth referred to as $\texttt{B13}$), and set $r_0 = 2000\mathrm{km}$. 
\begin{table*}
\centering
\begin{tabular}{l||l|r|r|r|r|r|r|r|r}
     Model & $\Omega_0\,[\rs]$& $B_0 \,[10^{12}\,\mathrm{G}]$ & $M_\mathrm{ZAMS}\,[\M]$ & $t_{\mathrm{pb}}$ [ms] & $t_{\mathrm{b}}$ [ms]& $P_{\mathrm{pre}}$[s] & $E_{\mathrm{GW}}\,[10^{-9}\,\M c^2]$ &$v_\mathrm{sh}^{\mathrm{max}}[\mathrm{km\,s}^{-1}]$ &$v_\mathrm{jet}^{\mathrm{avg}}[\mathrm{km\,s}^{-1}]$\\
    \hline
    \texttt{35M\_R0B12}  & 0   & 1 & 35 & 175 & 387.29 & - & 0.027 & 5000 & -
    \\
    \texttt{35M\_R07B12} & 0.7 & 1 & 35 & 200 & 388.37 & 8.98 & 0.142  & 7800 & -
    \\
    \texttt{35M\_R14B12} & 1.4 & 1 & 35 & 200 & 391.58 & 4.49 & 1.428 & 7500 &  4300
    \\
    \texttt{35M\_R21B12} & 2.1 & 1 & 35 & 175 & 397.00 & 2.99 & 6.407 & 9700 & 5900
    \\
    \texttt{35M\_R28B12} & 2.8 & 1 & 35 & 169 & 404.73 & 2.24 & 15.791 & 14500 & 9200
    \\
    \texttt{35M\_R35B12} & 3.5 & 1 & 35 & 155 & 414.93 & 1.80 & 21.90 & 21500 & 12200
    \\
    \texttt{35M\_R14B13} & 1.4 & 10 & 35 & 99 & 477.19 & 4.49 & 4.5054 & 15000 & 7800
    \\
    \texttt{35M\_R28B13} & 2.8 & 10 & 35 & 50 & 495.52 & 2.24 & 9.65 & 43200& 30300
    \\
    \texttt{25M\_R14B12} & 1.4 & 1 & 25 & 200 & 340.25 & 4.49 & 0.557 & 11500 & 5800
    \\
    \texttt{25M\_R28B12} & 2.8 & 1 & 25 & 190 & 353.05 & 2.24 & 8.294 & 33000 & 13800
    \\
\end{tabular}
\caption{Summary of models simulated. From left to right, we show the rotation rate ($\Omega_0$), seed magnetic field strength ($B_0$), progenitor mass ($M_\mathrm{ZAMS}$), duration of the simulation ($t_\mathrm{pb}$), pre-collapse period ($P_\mathrm{pre}$), energy emitted in GWs ($E_\mathrm{GW}$), maximal shock velocity $v_\mathrm{sh}^\mathrm{max}$ (rounded to the nearest hundred), and the average velocity of the shocked material after the jet is launched $v_\mathrm{jet}^\mathrm{avg}$ for all ten models (rounded to the nearest hundred). Note that no jet is launched for models \texttt{35M\_R0B12} and \texttt{35M\_R07B12}.}
\label{table:model_details}
\end{table*}

\section{Results}
\label{sec:results}

We simulate 10 pre-supernova models (see Section~\ref{sec:numerics} and Table~\ref{table:model_details}) and analyze the resulting explosion morphology, dynamics, and GW waveforms. During gravitational collapse, the central density rises rapidly to nuclear densities near $10^{14}\g\;\cm^{-3}$ after which it saturates and only gradually increases further. We define core bounce as the time when the central density reaches $8\times10^{14}\g\;\cm^{-3}$. Table~\ref{table:model_details} lists the bounce times $t_\mathrm{b}$ (in milliseconds) for the various models. The bounce time shows variation with progenitor mass, rotation profile, and magnetic field strength. In general, models with slower rotation reach core bounce at an earlier time than those with more rapid rotation.  The $25\M$ progenitors experience bounce slightly earlier than the $35\M$ models, while the two models with an initial field of $10^{13}\mathrm{G}$ exhibit significantly later bounce times (later by $\sim100$ms) than their $\texttt{B12}$ counterparts. In these $\texttt{B13}$  models, the time until bounce is prolonged by the higher magnetic pressure, which provides support against collapse \citep{Mösta2018}. While the $\texttt{B13}$ progenitors constitute an edge case, in the absence of observational or evolutionary constraints on the core magnetic fields in massive stars, it is worthwhile to investigate such extreme scenarios.

\subsection{Post-bounce Dynamics}\label{sec:pb_dyn}

All models with $\Omega_0 > 0.7 \rs$ result in a successful explosion, though their ejecta dynamics and morphology differ significantly.
\begin{figure}
    
    \includegraphics[width = 0.45 \textwidth]{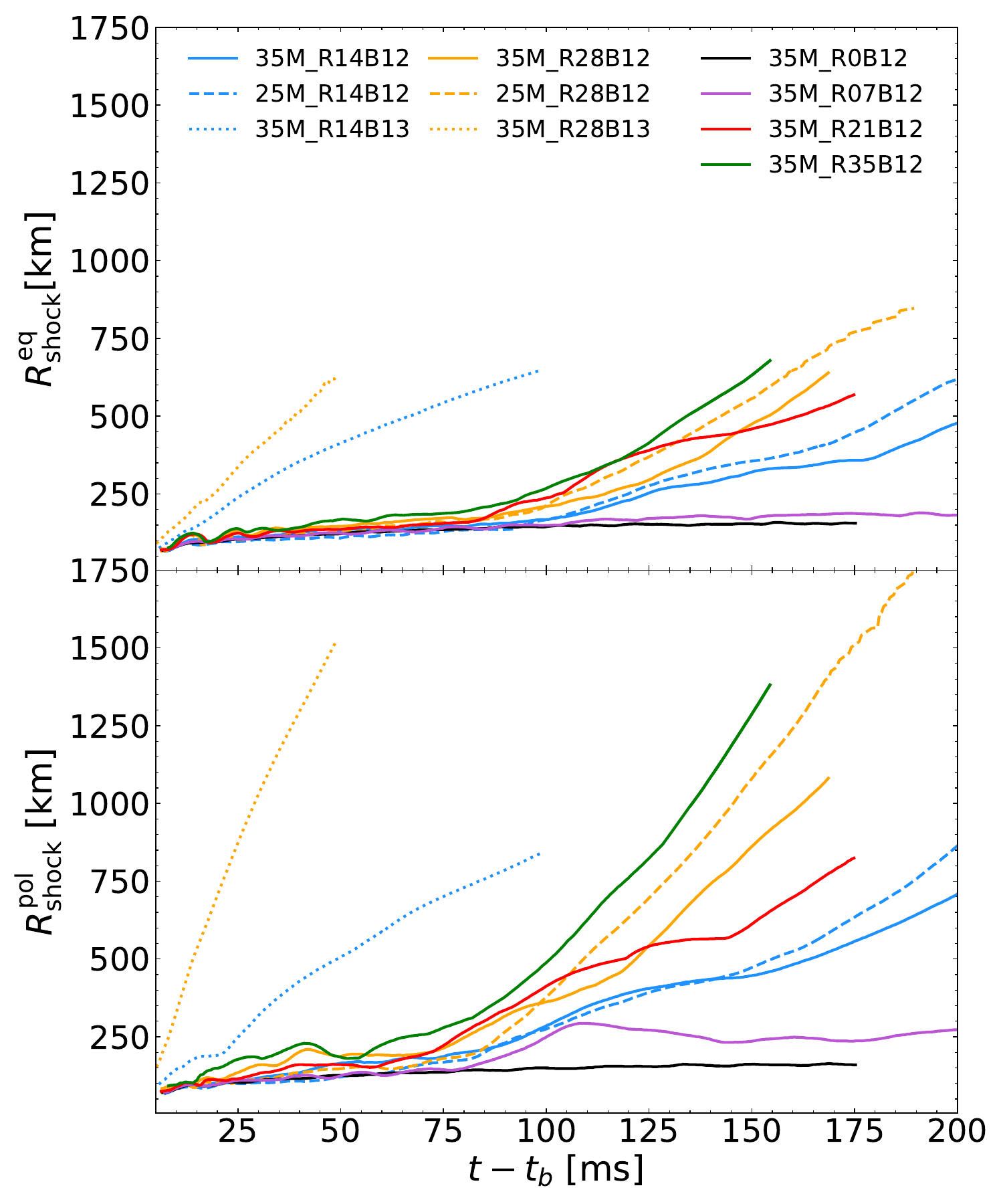}
    \caption{The top panel shows the average shock radius in the equatorial direction. The lower panel shows the shock radius in the polar direction as an average between the $\pm z$ for all ten models. Note that the shock radii for the $10^{13}\mathrm{G}$ models expand almost immediately after bounce, compared to the $\texttt{B12}$ progenitors.} 
    \label{fig:shock_radius}
\end{figure}
Fig.~\ref{fig:shock_radius} shows the average shock radius in the equatorial and polar and polar directions for the 10 models, calculated as the mean between the $+z$ and $-z$ directions. Initially, until about $60\mathrm{ms}$ post-bounce, all $\texttt{B12}$ models maintain equatorial and polar shock radii below $\sim200 \mathrm{km}$ indicating that shock remains approximately spherically symmetric. During this phase, the ongoing rotation continues to amplify the magnetic field via winding, ultimately enabling the formation of a strong toroidal component (as discussed in Section~\ref{sec:intro}). The magnetic pressure build-up then facilitates the eventual breakout of the bipolar jet. Beyond this point, for all of models with $\Omega_0> 0.7$ a jet is launched and the shock wave continues to propagate further outward. The ejected material of these explosion thus departs from axisymmetry, which in part linked to the $m=1$ kink instability \citep{Kink_instab_1991, Kink_instability_2014} causing the deformation of the jet.
Progenitors with a higher initial rotation rate, namely models \texttt{35M\_R21B12}, \texttt{35M\_R28B12}, \texttt{35M\_R35B12}, and \texttt{25M\_R28B12}, exhibit significantly more rapid shock expansion after approximately $100\mathrm{ms}$ post-bounce compared to the more slowly rotating $1.4\rs$ models (see Table~\ref{table:model_details} for the maximal shock and maximal jet velocities) resulting in more dramatically increasing polar and equatorial shock radii. The faster initial rotation leads to more rapid winding up of the field lines along the rotational axis, which then in turn leads to the jet being launched earlier than for more slowly rotating models. This is also consistent with the maximal shock velocities observed for each model (see Table~\ref{table:model_details}): more rapid rotation is related to more rapid and more dramatic shock expansion. In contrast, progenitors \texttt{35M\_R0B12} and \texttt{35M\_R07B12} show very little to no shock expansion during the simulation period. This suggests that the energy accumulated behind the stalled shock is insufficient for shock revival, leading to a failed explosion for these progenitors. 

Both highly-magnetized $\texttt{B13}$ models show very rapid shock expansion immediately after core bounce. The $\texttt{35M\_R28B13}$ progenitor exhibits very high (polar) shock velocities ($\sim43000\mathrm{km}\;\s^{-1}$) almost immediately after core bounce. However, although the shock velocities attained by this model are high, they are within the range of what is expected for magento-rotational CCSNe \citep{Burrows2007}.  Overall, the $25\M$ models with a $\texttt{B12}$ magnetic field exhibit more pronounced shock expansion than the more massive $35\M$ progenitors. As these progenitors differ in properties beyond mass, we cannot isolate the impact of specific progenitor traits on the shock evolution and morphology. Nevertheless, it is well established that progenitor structure can significantly impact the resulting supernova explosion (e.g. \cite{Naveen2020}), highlighting the need for further systematic studies in the future. Finally, we wish to note that the wiggle observed in the shock radius of model \texttt{25M\_R28B12} beginning around $170\mathrm{ms}$ is attributed to a change in the refinement level - from level 6 to level 5 - used to track the expanding shock.
\begin{figure*}
    \centering
    \includegraphics[width =  \textwidth]{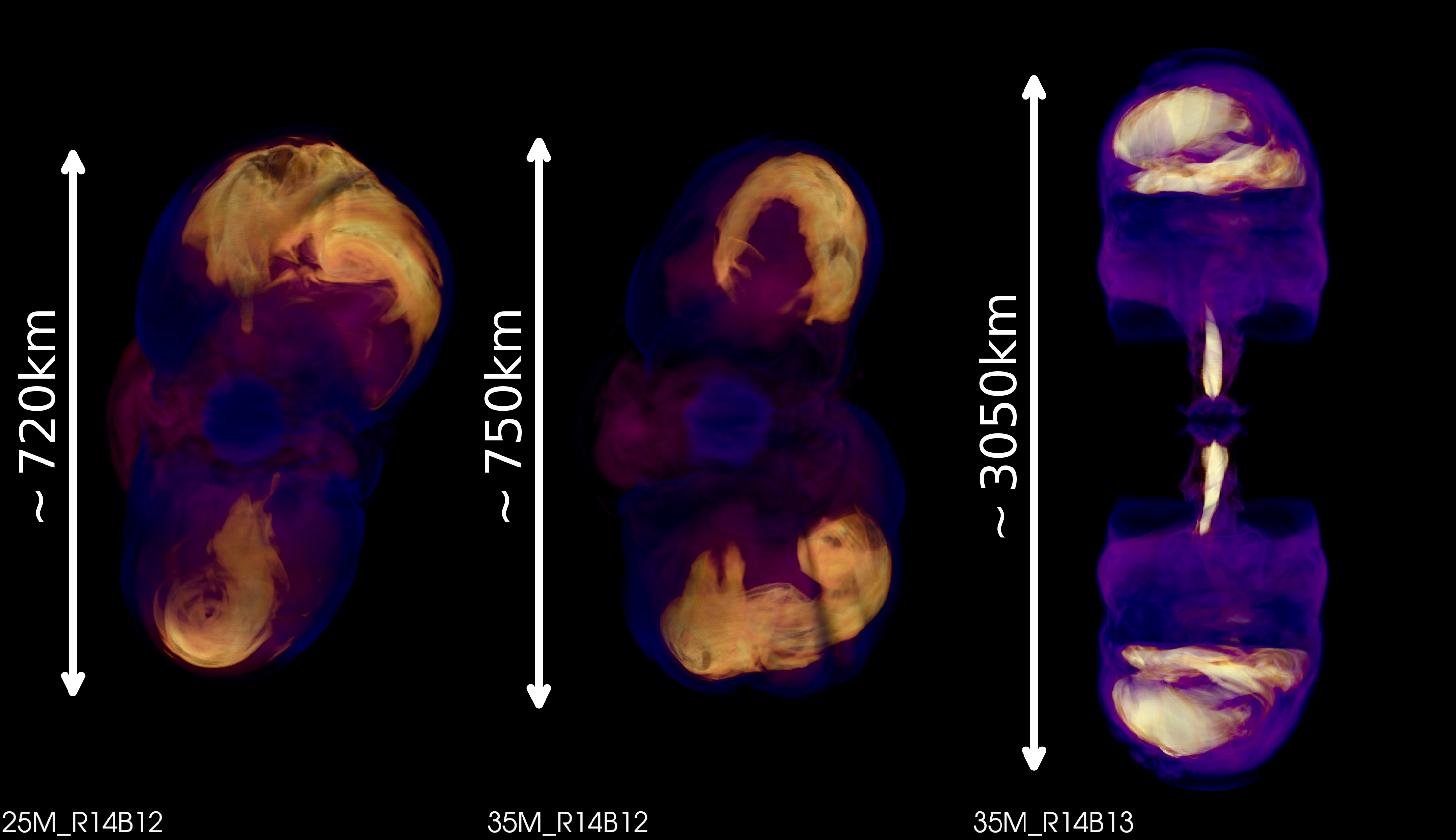}
    \caption{Volume renderings of specific entropy as viewed from the $yz$-plane for progenitors $\texttt{25M\_R28B12}$ and $\texttt{35M\_R28B12}$ at $100\mathrm{ms}$ post bounce, and $\texttt{35M\_R28B13}$ at $\sim50\mathrm{ms}$ post bounce (not to scale). The shock radius for the progenitors with a $\texttt{B12}$ field is similar and spans $\sim720\mathrm{km}$ and $\sim750\mathrm{km}$ respectively. The shock for the $\texttt{B13}$ progenitor is much more extensive, spanning $\sim3050\mathrm{km}$.}
    \label{fig:R28_VolumeRenderings}
\end{figure*}

In Fig.~\ref{fig:R28_VolumeRenderings} we show volume renderings of the entropy for the $2.8\rs$ progenitors. Here, the $\texttt{B13}$ model has a higher entropy near the rotation axis and in the polar ejecta and also exhibits a more collimated jet compared to the $\texttt{B12}$ progenitors with the same rotation rate. The morphology of the $\texttt{B13}$ model's shock is more similar to the 2D model in \citet{Moesta_E25, Burrows2007}, which is associated with the significantly stronger initial poloidal field, which stabilizes the jet against deformation by a kink instability \citep{Mösta2018}. In all three progenitor models we show in Fig.~\ref{fig:R28_VolumeRenderings}, the regions of highest entropy are in the ejecta lobes and near the axis of rotation, indicating strong shock heating in this region. 
\begin{figure}
    \centering
    \includegraphics[width = 0.45\textwidth]{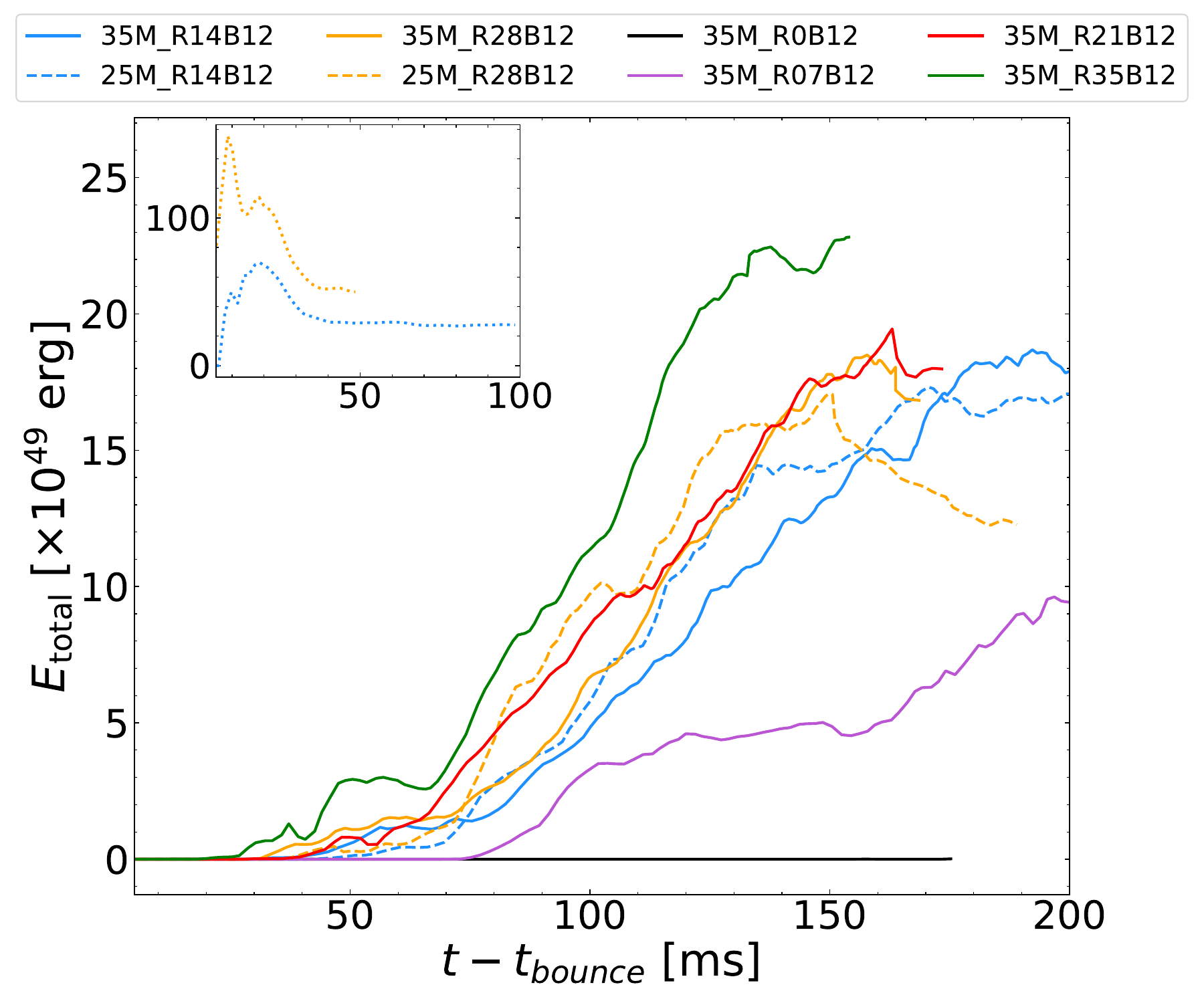}
    \caption{The explosion energy as a function of time for all the eight models with a $10^{12}\mathrm{G}$ seed magnetic field is shown in the main plot. The inset shows the explosion energy of the $10^{13}\mathrm{G}$ progenitors.}
    \label{fig:ej_nrg}
\end{figure}

Fig. \ref{fig:ej_nrg} shows the total explosion energy - which is a sum of the kinetic, magnetic, and internal energies - for the 10 models. Note that the $\texttt{B13}$ models - shown in the inset for clarity of presentation - exhibit explosion energies an order of magnitude higher than those with a $\texttt{B12}$ seed field. The rotating $\texttt{B12}$ models' energy increases monotonically with time for the duration of our simulation with the exception of the $\texttt{35M\_R28B12}$ model which reaches a maximum near 150ms and begins falling off after. The most rapidly rotating models exhibit higher explosion energies compared to the progenitors with a lower initial rotation rate. Note that the explosion energy for the non-rotating model is zero throughout, as this model did not explode successfully. The $\texttt{B13}$ models show a rapid increase of explosion energy immediately after bounce and then fall-off again, plateauing to $35\times10^{49}$erg for \texttt{35M\_rot14B13} and $55\times10^{49}$erg for \texttt{35M\_rot28B13} after about $40\mathrm{ms}$ respectively. These results are in good agreement with what was found by \citep{Shibagaki_GWs_2024}.

\begin{figure}
     \centering
     \includegraphics[width = 0.45\textwidth]{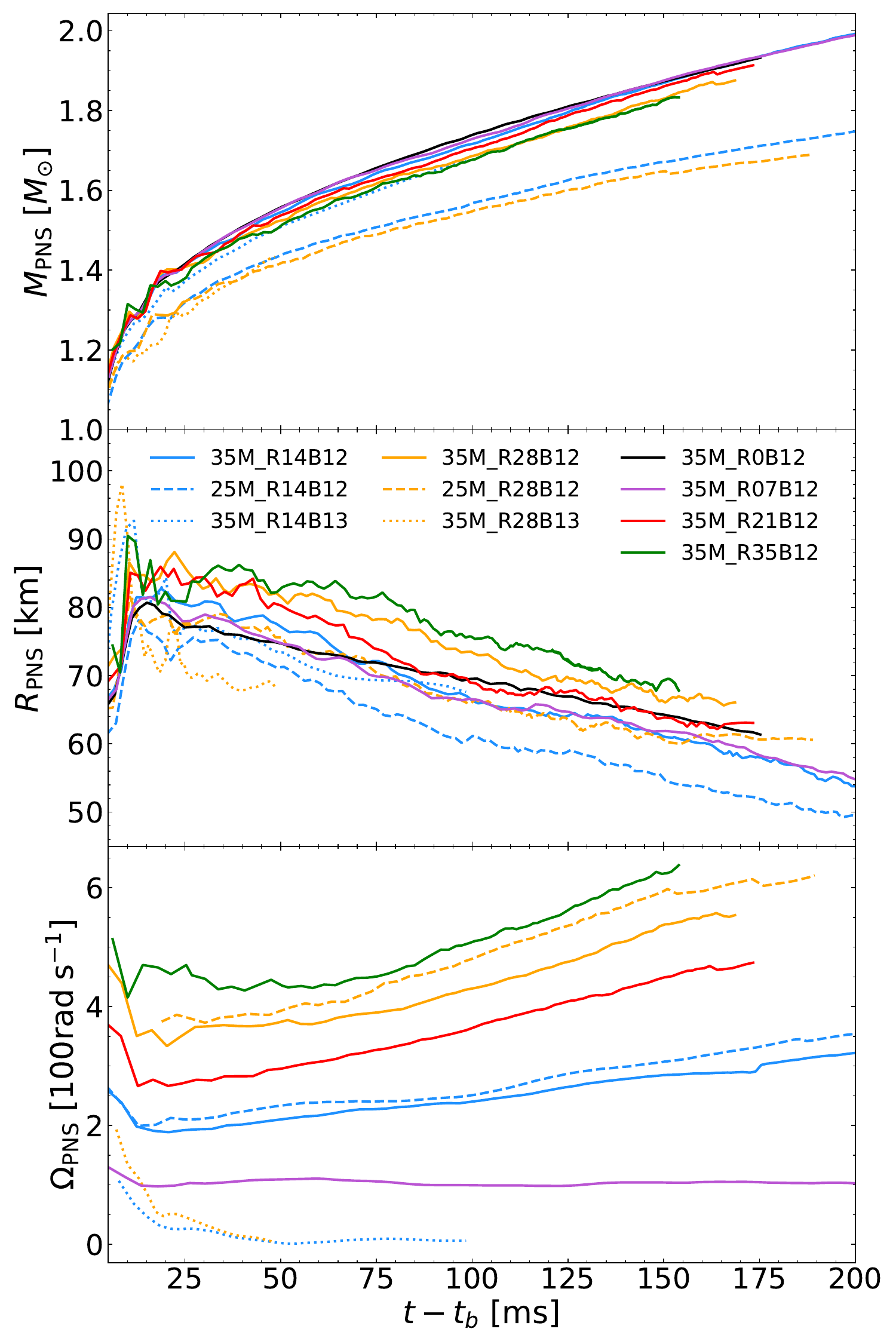}
     \caption{For all 10 models: In the top panel we show the PNS mass in $\M$, defined as the total mass enclosed within the iso-surface at which $\rho = 10^{11}\g\,\cm^{-3}$ and in the middle panel we show the PNS radius. For the 9 rotating models: In the bottom panel we show the PNS rotation rate.}
     \label{fig:PNS}
\end{figure}
In the top panel of Fig.~\ref{fig:PNS} we show the temporal evolution of the PNS mass in $\M$ for all progenitor models. We observe that the PNS mass evolves similarly for all $35\M$ models with only minor differences between the initial rotation rates. The PNS mass increases monotonically due to mass accretion and the $35\M$ models (except $\texttt{35M\_R28B13}$) all form a PNS with a mass between 1.8 and 2.0 $\M$ after 200 ms. This agrees with the current range and upper limits on the maximal (baryonic) mass for PNSs (see for example \citep{OConnor2011, ObergaulingerAloy2021}). The $25\M$ progenitors result in a slightly less massive PNS of $\sim1.7\M$.
The middle panel of Fig.~\ref{fig:PNS} shows the evolution of the PNS radius in $\mathrm{km}$ which decreases monotonically as the PNS core density increases. Here we observe that the impact of the progenitor star's rotation is more significant than for the PNS mass evolution. Finally, in the bottom panel of Fig.~\ref{fig:PNS} we show the average PNS rotation rates for our 9 rotating models. The average rotation rate for all $\texttt{B12}$ models continuously increases during the simulated time due to ongoing accretion and angular momentum transport from the surrounding stellar material. In contrast, the $\texttt{B13}$ models experience enhanced magnetic braking due to the strong magnetic fields which effectively extract angular momentum from the PNS.

\subsection{Gravitational Wave Signals}\label{sec:GWs}

In this Section we discuss the GW signatures resulting from matter motions of our 10 models and highlight the impact rotation and magnetic fields have on the signals. As this work uses an approximate neutrino treatment (see Section~\ref{sec:numerics}) we cannot consider GWs from anisotropic neutrino emission. We compute gravitational wave strains and amplitudes using the reduced mass-quadrupole tensor $I_{jk}$, which provides an adequate approximation for capturing gravitational wave emission in core-collapse supernovae \citep{QuadrupoleExtractionMethod}.
We follow the approach outlined by \citet{KurodaTaKo_14, h_char_Shibagaki2021} and \citet{ScheideggerKWF_10}. The GW strains are thereby given as
\begin{align}
    h_{+,\mathrm{eq}} = \dfrac{G}{c^4} \left(\ddot{I}_{zz} - \ddot{I}_{yy}\right),  \label{eqn:strains1}\\
    h_{\times,\mathrm{eq}} = \dfrac{2 G}{c^4} \left(\ddot{I}_{yz}\right), \label{eqn:strains2}\\
    h_{+,\mathrm{p}} = \dfrac{G}{c^4} \left(\ddot{I}_{xx} - \ddot{I}_{yy}\right), \label{eqn:strains3}\\
    h_{\times,\mathrm{p}} = \dfrac{-2 G}{c^4} \left(\ddot{I}_{xy}\right), \label{eqn:strains4}
\end{align}
where the subscripts \textit{p} and \textit{eq} refer polar and equatorial observer, respectively. Note that the above strains will have to be scaled by $1/D$ where $D$ is the distance to the source when computing the GW spectral energy density. The GW energy emitted is expressed as
\begin{align}
    E_\mathrm{GW} &= \dfrac{1}{5}\dfrac{G}{c^5} \int_{-\infty}^{\infty}dt \dddot I_{ij}\dddot I_{ij}. 
    \label{eqn:E_GW}
\end{align}
We compute the characteristic strain as described in \citet{h_char_Shibagaki2021}
\begin{align}
    h_\mathrm{char}(f) &= \sqrt{\dfrac{8}{\pi}\dfrac{G}{c^3}\dfrac{1}{D^2}\dfrac{dE_{GW}}{df}},
    \label{eqn:h_char}
\end{align}
where 
\begin{align}
    \dfrac{1}{D^2}\dfrac{dE_{GW}}{df} &=  \dfrac{\pi f^2}{2}\dfrac{c^3}{G} \left(|\tilde{h}_+(f)|^2+|\tilde{h}_\times(f)|^2\right), 
\end{align}
and where $\tilde{h}$ is the Fourier transform of the GW strain $h$. Note that $\dfrac{dE_{GW}}{df}$ is estimated as an angle average.
\begin{figure*}
    \centering
    \includegraphics[width =  \textwidth]{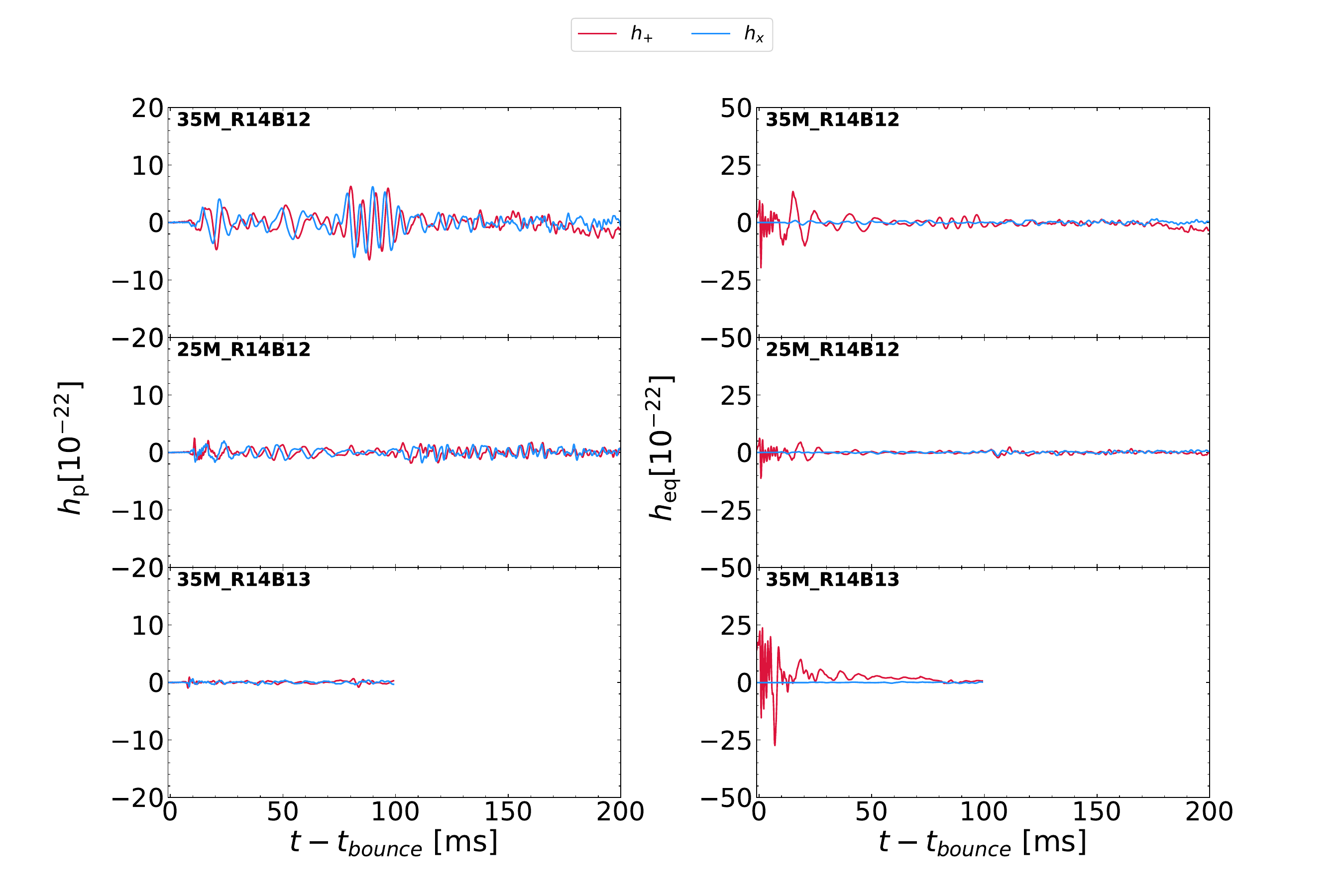}
    \caption{The (\(+\)) and (\(\times\)) polarizations of the gravitational-wave strain as seen by a polar (left) and equatorial (right) observer as a function of time for the progenitors with a rotation rate of $\Omega0=1.4\rs$. The amplitudes are scaled to a source distance of $10\,\rm{kpc}$.}
    \label{fig:GW_R14}
\end{figure*}

In Fig.~\ref{fig:GW_R14} we show the GW strain for the progenitors where $\Omega_0 = 1.4\rs$, scaled to a source distance of $10\mathrm{kpc}$. The right-hand column shows the GW component for an equatorial observer. At core bounce, all three models exhibit a pronounced bounce spike signal in the $h_+$ polarization. This is associated with the rapid in-fall of matter, which is abruptly halted at core bounce as the EOS stiffens, and then reverts its direction to become an outward shock wave on the time-scale of $\mathcal{O}(1\mathrm{ms})$. This spike is then followed by a short ring-down phase as the core settles back into equilibrium. The $25\M$ progenitor exhibits a slightly weaker signal than the $35\M$, with an amplitude only half as strong at bounce. The $\texttt{35M\_R14B13}$ model shows a strong bounce signal and post-bounce ring-down features, and the strain remains positive after bounce, in accordance with the behavior described in \citep{Shibata2006}: the GW amplitude increases after core-bounce due to MHD outflows. 

The left column of Fig.~\ref{fig:GW_R14} shows the polar component of the signal, for an observer located along the axis of rotation of the star. All three models exhibit some oscillatory features associated with convective matter motions and rotational instabilities during the more dynamical post-bounce phase where the jet is still forming and the stellar plasma sloshes around the newly formed PNS. For model \texttt{35M\_rot14B12} SASI-like activity develops between $\sim75-100\mathrm{ms}$ post-bounce. The $25\M$ progenitor exhibits a more subdued polar GW signature than the more massive $35\M$ one. Compared to the $\texttt{B12}$ models, however, the polar strain of the more strongly magnetized $\texttt{35M\_R14B13}$ progenitor is strongly suppressed. This suppression is likely associated with MHD effects on the material surrounding the PNS and the the nearly 2D, axisymmetric behavior of these models, which results in very little matter motions in the xy-plane. The strong poloidal field decelerates the matter in the xy-plane, reducing the growth of matter asymmetries, collimating the flows, and thereby suppressing the poloidal GW amplitudes \citep{ScheideggerKWF_10}. 

\begin{figure*}
    \centering
    \includegraphics[width =  \textwidth]{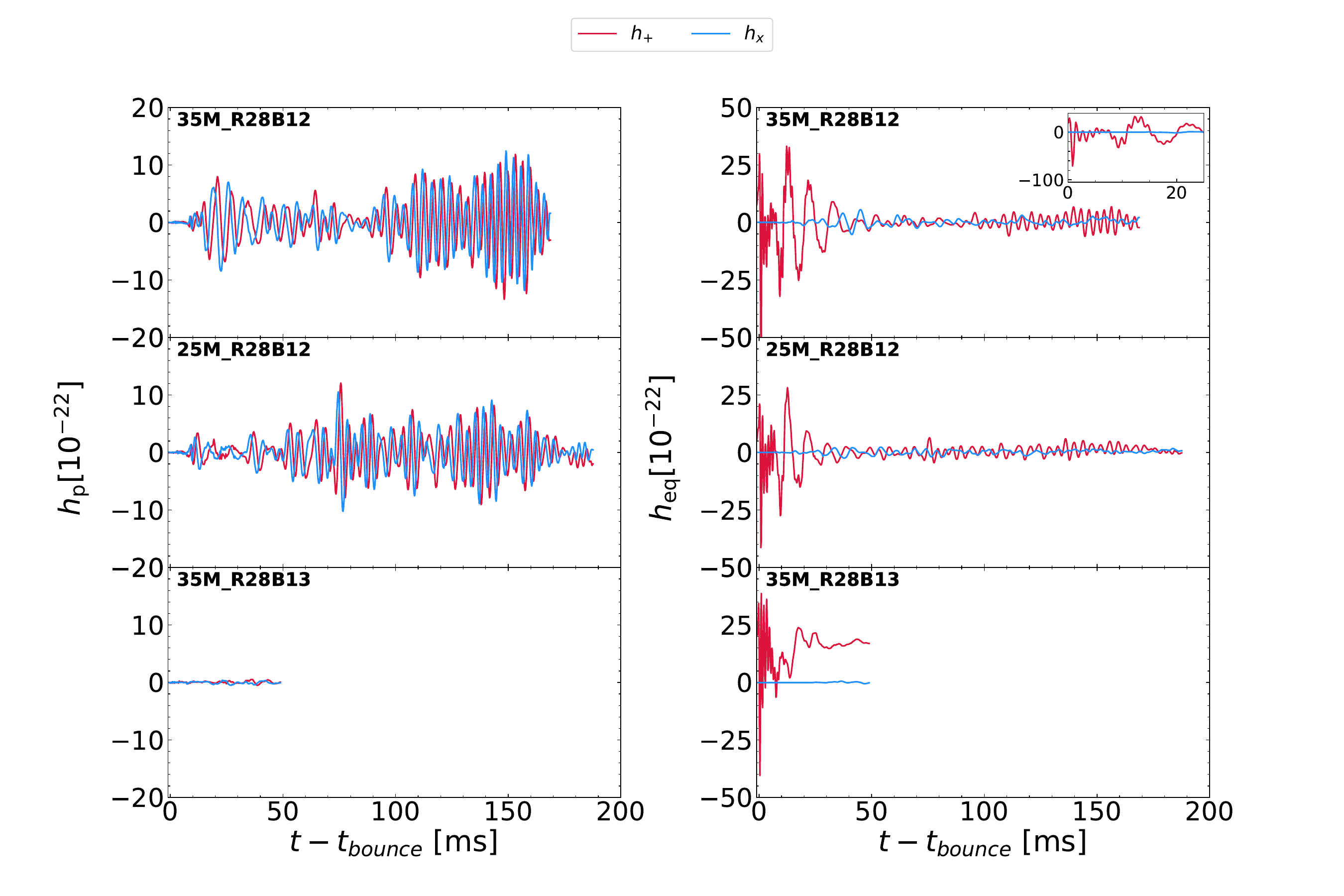}
    \caption{The (\(+\)) and (\(\times\)) polarizations of the gravitational wave strain as seen by an polar (left) and equatorial (right) observer are shown as a function of time for the progenitors with a rotation rate of $2.8\rs$. The amplitudes are scaled to a source distance of 10 kiloparsecs.}
    \label{fig:GW_R28}
\end{figure*}
As described above for the  $\Omega_0 = 1.4\rs$ models, all three progenitors with $\Omega_0 = 2.8\rs$ in Fig.~\ref{fig:GW_R28}, exhibit a bounce spike in the equatorial plane. The $35\M$ model's signal is weaker than than that of the $25\M$ progenitor, with a bounce spike only approximately half the strength of the more massive star. The strong magnetic field of $\texttt{35M\_R28B13}$ also suppresses the bounce signal as explained above. In this case, the $\texttt{B13}$ model does not show the typical ring-down phase, but instead remains positive post-bounce, plateauing between $5\mathrm{ms}$ and the end of the simulation in accordance with \citep{Shibata2006}.

In the polar plane (left side of Fig.~\ref{fig:GW_R28}), the $\texttt{B12}$ models exhibit strong post-bounce oscillations exceeding the amplitudes attained by the respective $\Omega_0 = 1.4\rs$ models. Before the jet is launched ($\sim75\mathrm{ms}$) the shocks in progenitors $\texttt{35M\_R28B12}$ and $\texttt{25M\_R28B12}$ are disturbed only by convective flows and the growth of large-scale bubbles. SASI-like activity becomes apparent until $\sim100\mathrm{ms}$. As before, the $\texttt{B13}$ model has a strongly suppressed polar signal.

\begin{figure*}
    \centering
    \includegraphics[width =  \textwidth]{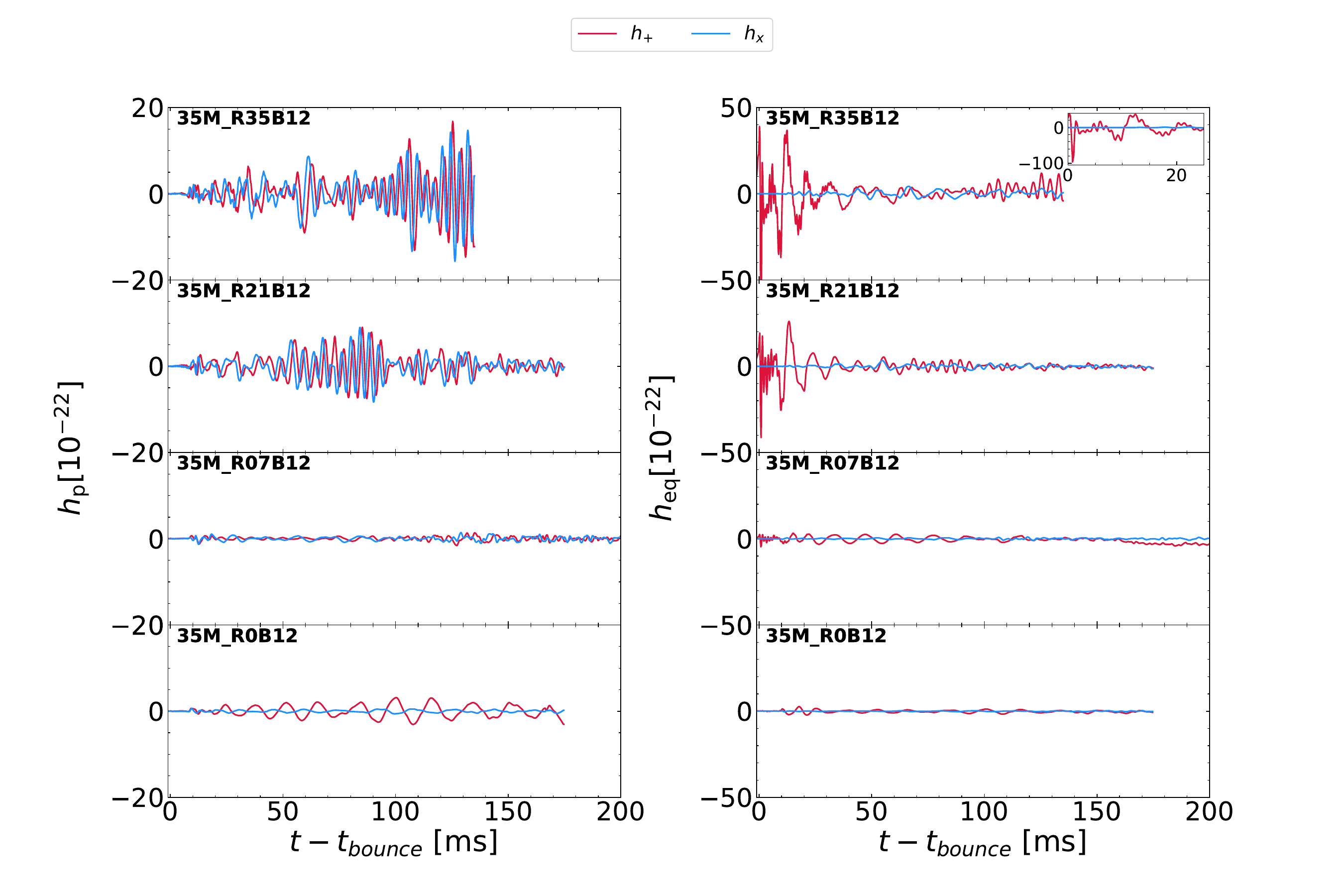}
    \caption{The (\(+\)) and (\(\times\)) polarizations of the gravitational wave strain as seen by a polar (left) and equatorial (right) observer as a function of time for the progenitors with a rotation rate of $0, 0.7, 2.1$, and $3.5\rs$. The amplitudes are scaled to a source distance of 10 kiloparsecs.}
    \label{fig:GW_Rest}
\end{figure*}
Finally, in Fig.~\ref{fig:GW_Rest} we compare the remaining four progenitors $\texttt{35M\_R0B12}$, $\texttt{35M\_R07B12}$, $\texttt{35M\_R21B12}$, and $\texttt{35M\_R35B12}$. The equatorial strain in $h_+$ allows us to see that there is a clear correlation between the strength of the bounce spike and the rotation rate: the more rapidly the progenitor star rotates, the more pronounced the resulting spike in the equatorial plane at bounce will become. Out of our 9 rotating models, the $\texttt{35M\_R07B12}$ model exhibits the lowest amplitude signal, whereas the $\texttt{35M\_R35B12}$ model exhibits the most dramatic GW signature. A more rapid rotation rate is associated with a more rotationally flattened inner core at bounce and thus results in a more strongly pronounced initial bounce spike followed by a more dramatic ring-down phase. Contrary to the rotating models, however, $\texttt{35M\_R0B12}$ does not exhibit any bounce signal, as the collapse of the core is highly symmetric. All GW emission in this case is likely associated with fluid motions around the accreting PNS and prompt convection. 

For a polar observer (left side of Fig.~\ref{fig:GW_Rest}) more rapid rotation leads to oscillations of somewhat higher amplitude than models with lower rotation, associated with deformations of the matter flows in the xy-plane. However, the morphology between the polar signals is quite different from model to model. The $\texttt{35M\_R35B12}$ progenitor launches a jet around $\sim 75\mathrm{ms}$ (See Fig.~\ref{fig:shock_radius}), beyond $\sim 90\mathrm{ms}$ signs of SASI activity and convective plasma motions in the accreting matter become evident. Similarly, the $\texttt{35M\_R21B12}$ progenitor launches a jet around $\sim 70 \mathrm{ms}$, followed by a phase of SASI-like activity and convective flows. The non-rotating model exhibits oscillatory features starting around $10\mathrm{ms}$ post bounce in the $h_+$. This is associated with motions of the stellar plasma around the newly-formed PNS, stretching and compressing along the x- and y-axis, as seen in the almost sinusoidal $h_+$ signal.

In Table~\ref{table:model_details} we summarize the maximum GW energy attained during the simulation time for each model. The energies span four orders of magnitude between the non-rotating and the most-rapidly rotating models. The influence of the rotation rate (and progenitor mass) on the amount of gravitational energy released is evident here as well: more rapidly rotating models exhibit higher energies than those initialized with a lower $\Omega_0$.  The impact of the magnetic field is also evident when comparing the $\Omega_0 = 1.4\rs$ and $\Omega_0 = 2.8\rs$ models to their $\texttt{B13}$ configurations. In both cases a stronger magnetic field results in significantly higher gravitational wave energy (despite the fact that the $\texttt{B13}$ models were only simulated for a fraction of the total post bounce time). Finally, the $25\M$ models' GW energies are only approximately half of the energy emitted by the $35\M$ progenitors with the same rotation profile and magnetic field strength. All rotating models exhibit a steep growth in the gravitational-wave energy in the first $10\mathrm{ms}$ associated with turbulent motions at core bounce, the stalling shock, and initial accretion onto the PNS. A second increase is evident for some of the models near near the time where a jet is launched. This is linked to a period of increased activity in the polar GW strain seen in the left panels of Fig.~\ref{fig:GW_R14}-\ref{fig:GW_Rest}.

\begin{figure*}
  \centering
  
  \centerline{\textbf{$M_\mathrm{ZAMS} = 25\M, \;\;$  $\Omega_0 = 1.4\rs$}}
  \vspace{0.2cm}
  \begin{subfigure}[t]{0.42\textwidth}
    \includegraphics[width=\linewidth]{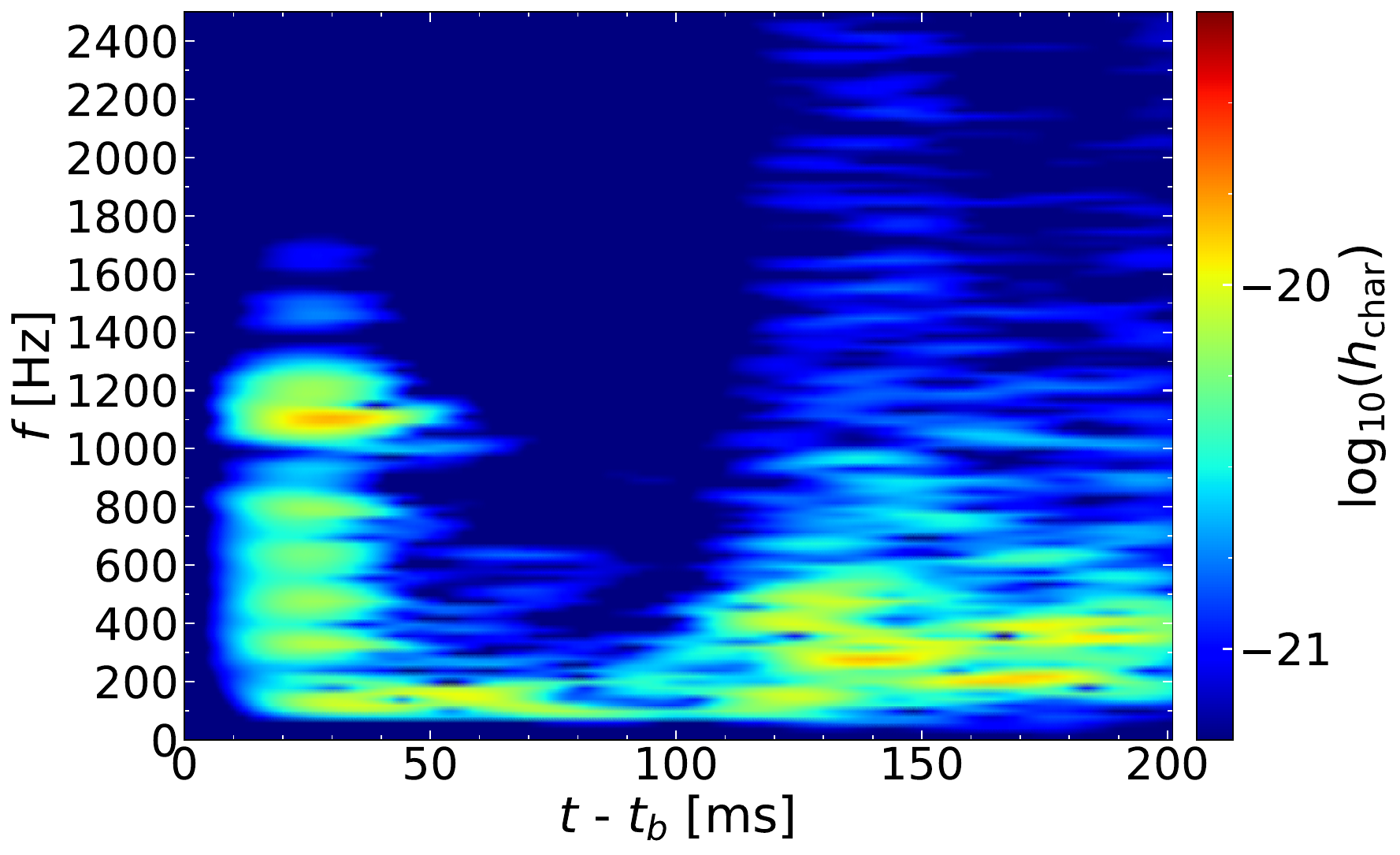}
  \end{subfigure}\hfill
  \begin{subfigure}[t]{0.42\textwidth}
    \includegraphics[width=\linewidth]{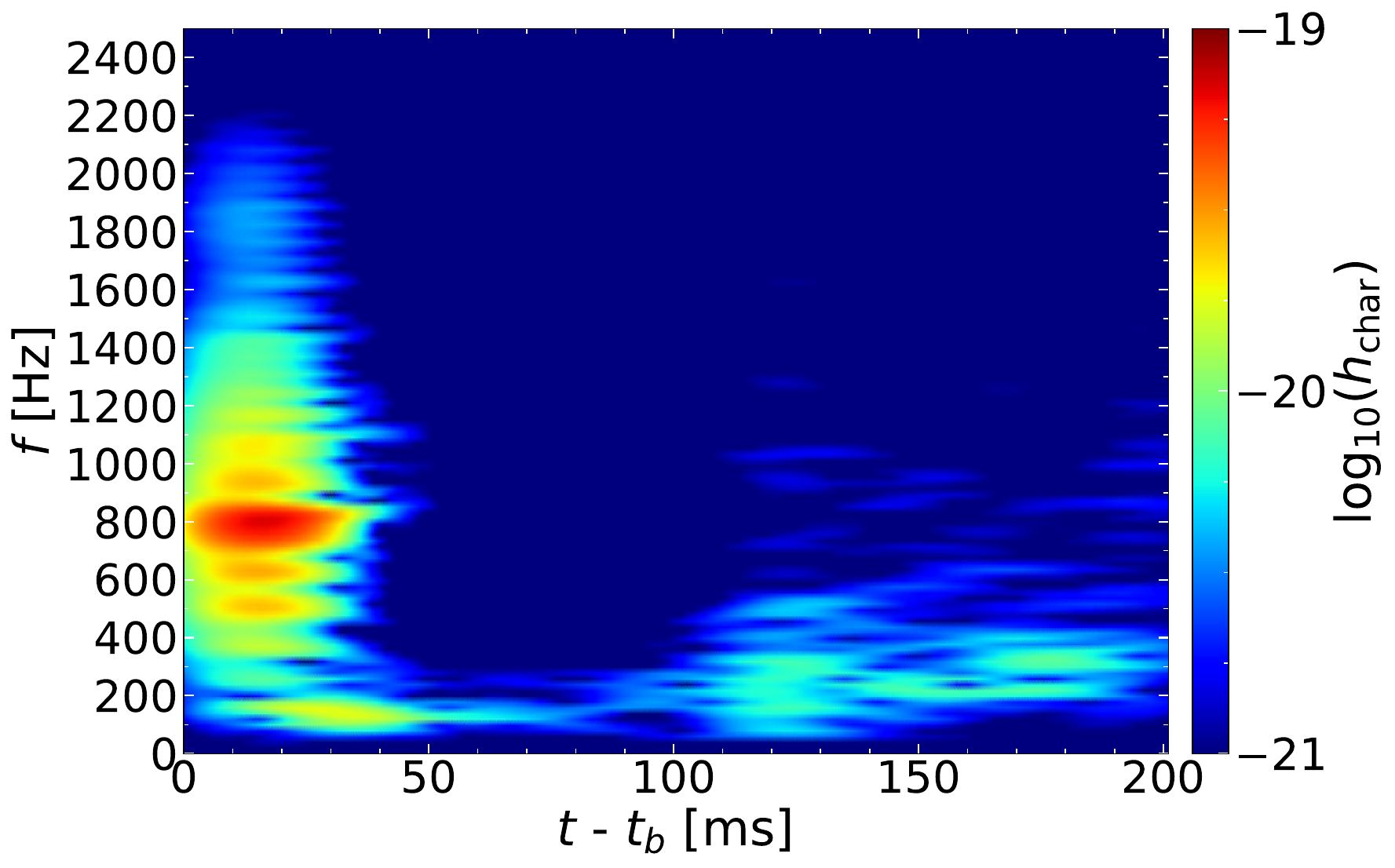}
  \end{subfigure}

  \centerline{\textbf{$M_\mathrm{ZAMS} = 35\M, \;\;$  $\Omega_0 = 1.4\rs$}}
  \vspace{0.2cm}
  \begin{subfigure}[t]{0.42\textwidth}
    \includegraphics[width=\linewidth]{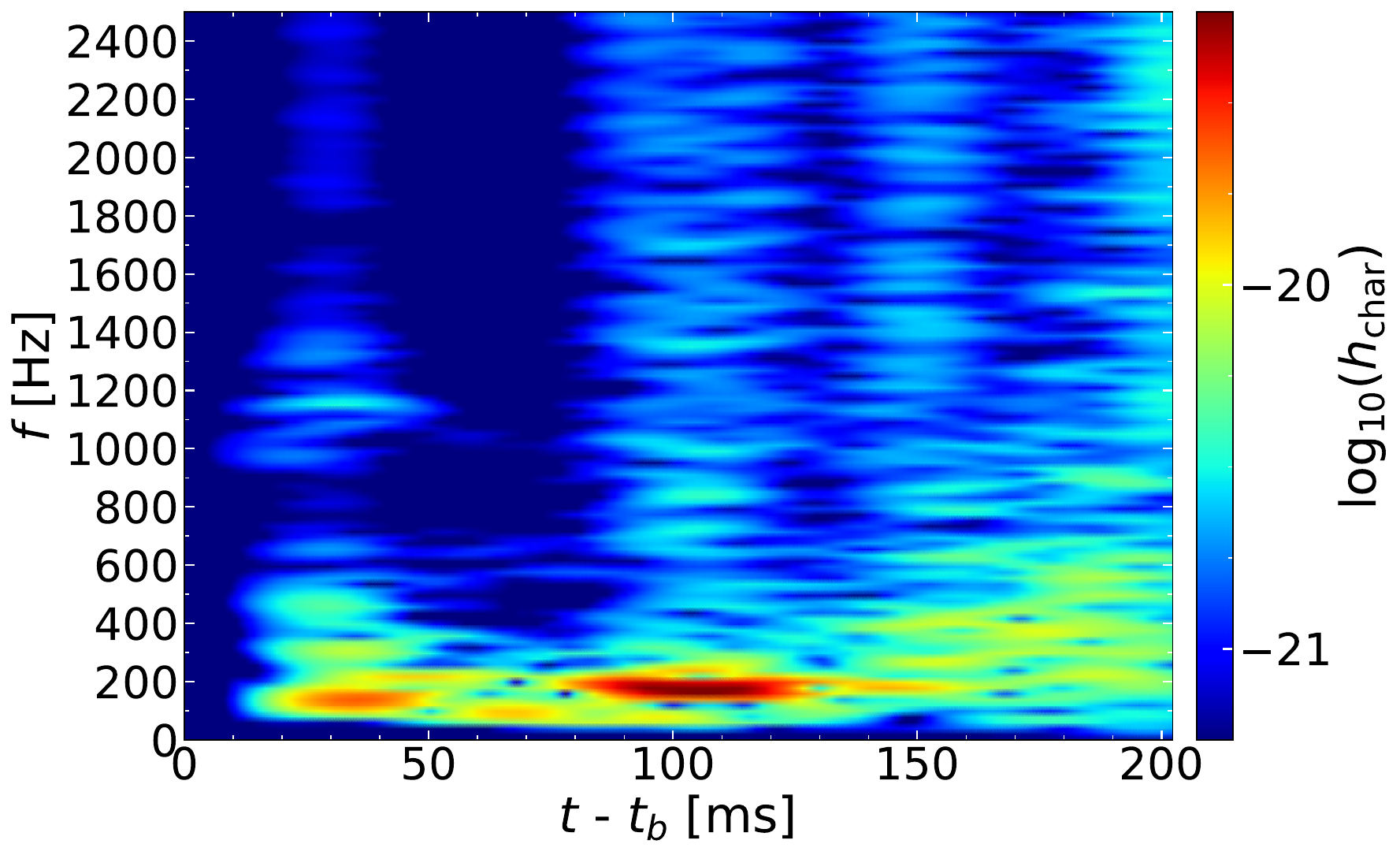}
  \end{subfigure}\hfill
  \begin{subfigure}[t]{0.42\textwidth}
    \includegraphics[width=\linewidth]{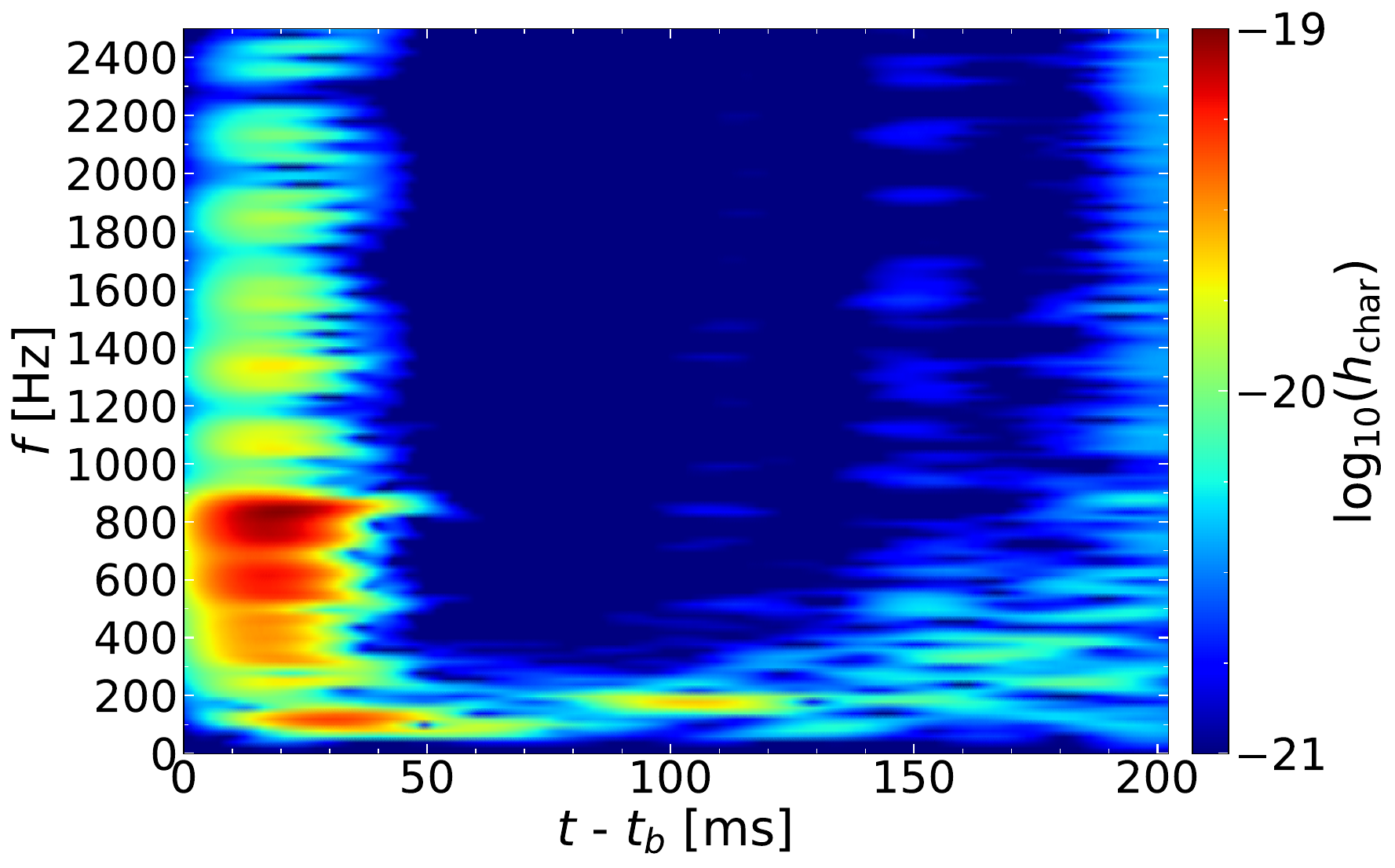}
  \end{subfigure}

  \centerline{\textbf{$M_\mathrm{ZAMS} = 25\M, \;\;$  $\Omega_0 = 2.8\rs$}}
  \vspace{0.2cm}
  \begin{subfigure}[t]{0.42\textwidth}
    \includegraphics[width=\linewidth]{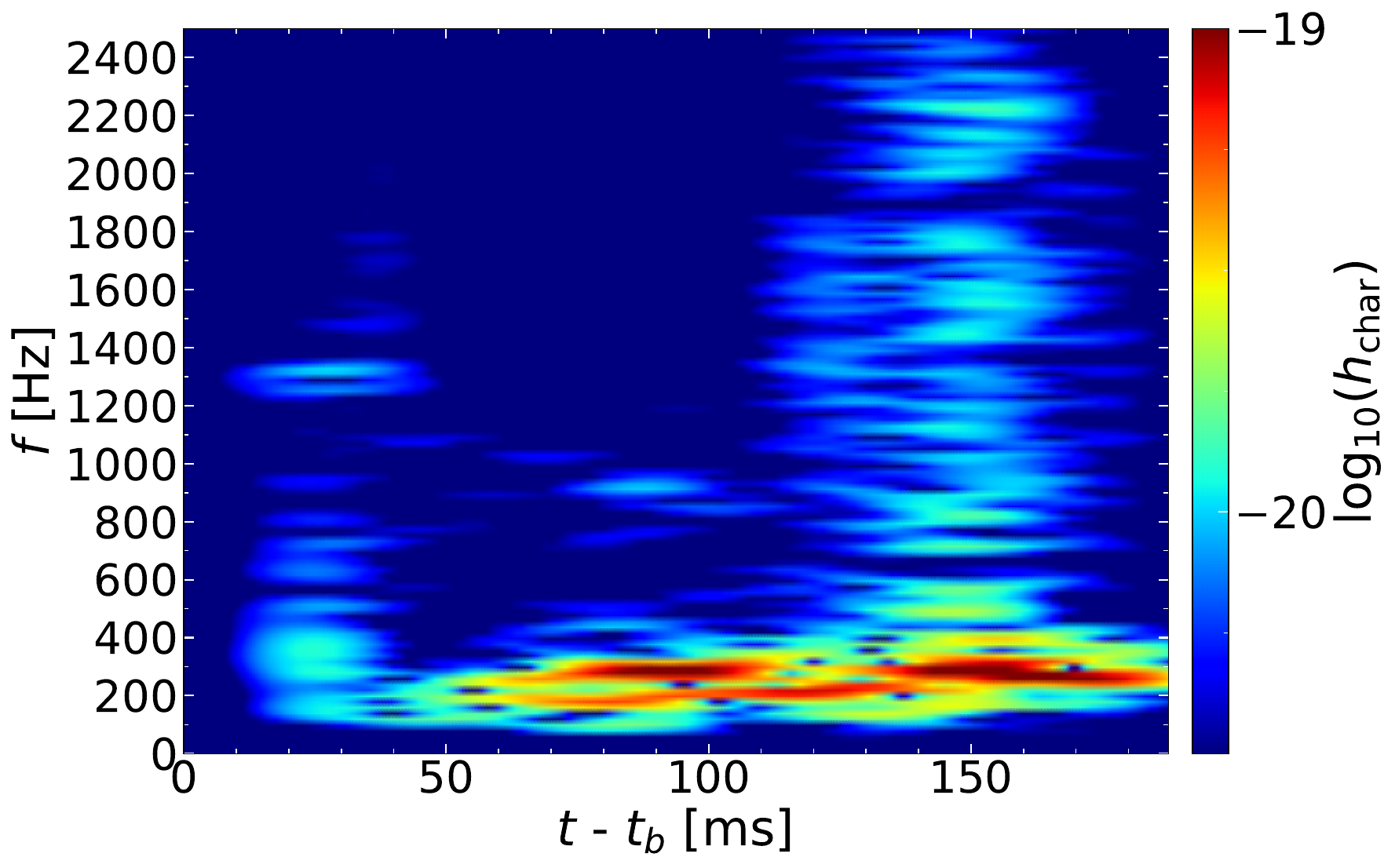}
  \end{subfigure}\hfill
  \begin{subfigure}[t]{0.42\textwidth}
    \includegraphics[width=\linewidth]{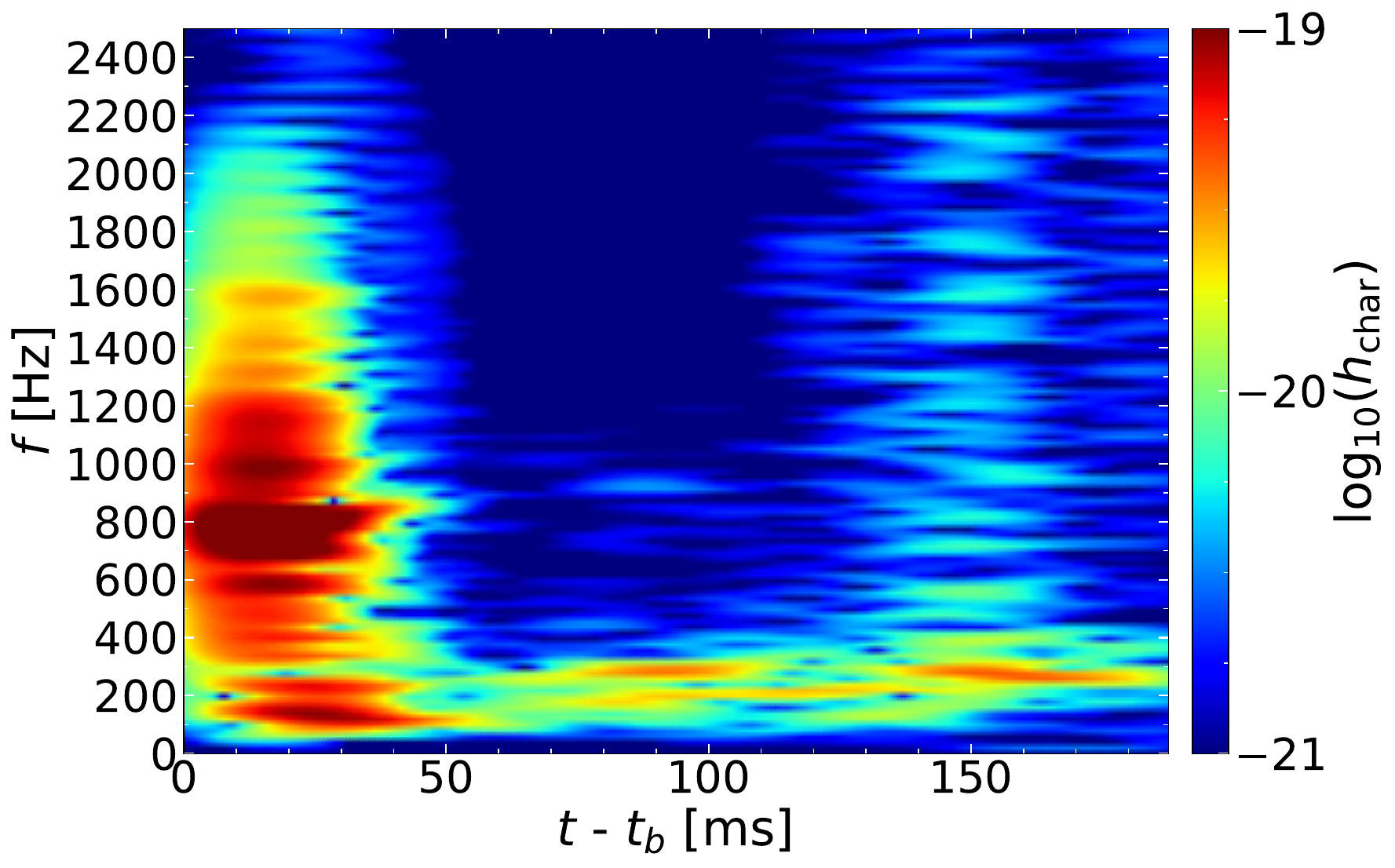}
  \end{subfigure}

  \centerline{\textbf{$M_\mathrm{ZAMS} = 35\M, \;\;$  $\Omega_0 = 2.8\rs$}}
  \vspace{0.2cm}
  \begin{subfigure}[t]{0.42\textwidth}
    \includegraphics[width=\linewidth]{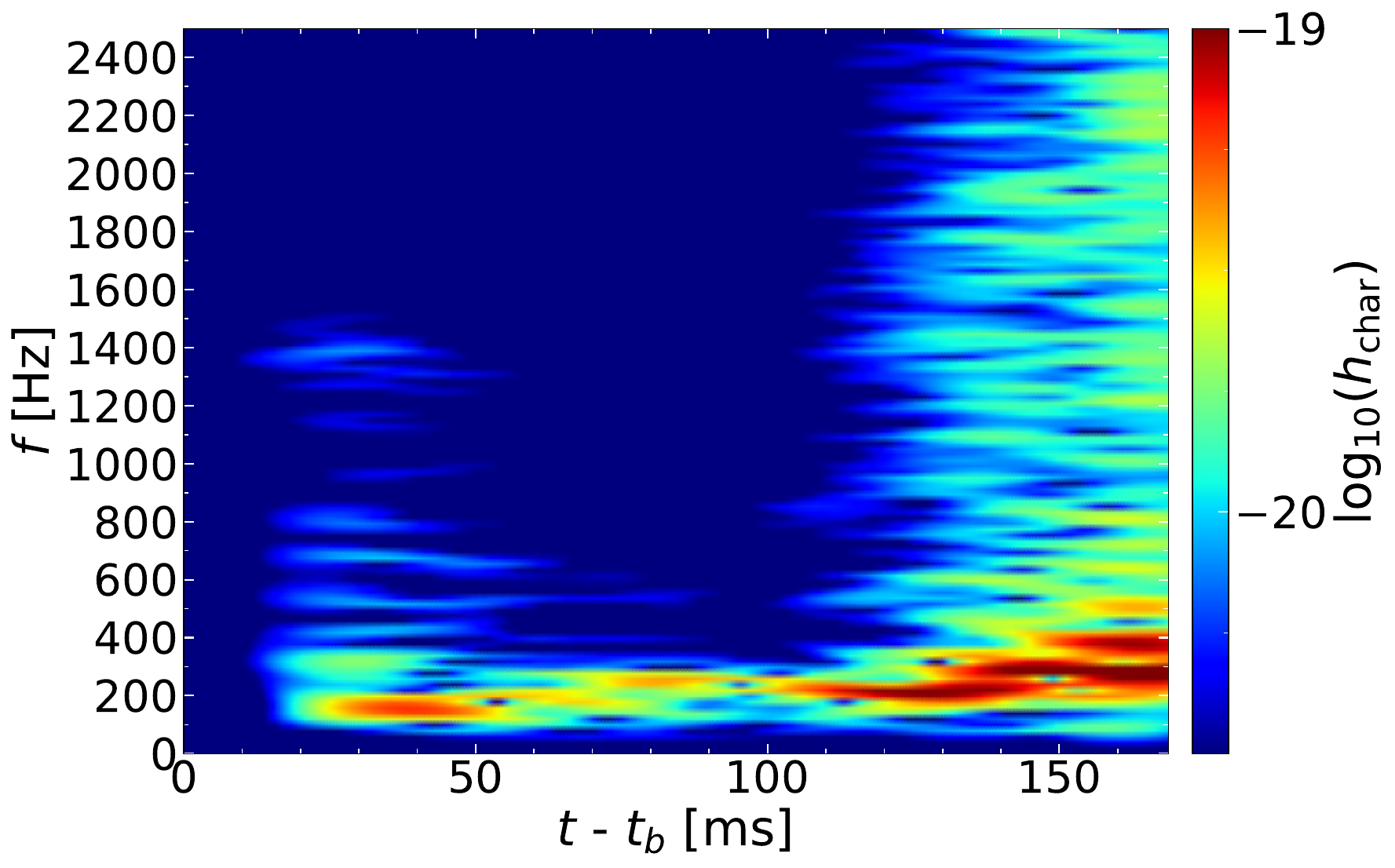}
  \end{subfigure}\hfill
  \begin{subfigure}[t]{0.42\textwidth}
    \includegraphics[width=\linewidth]{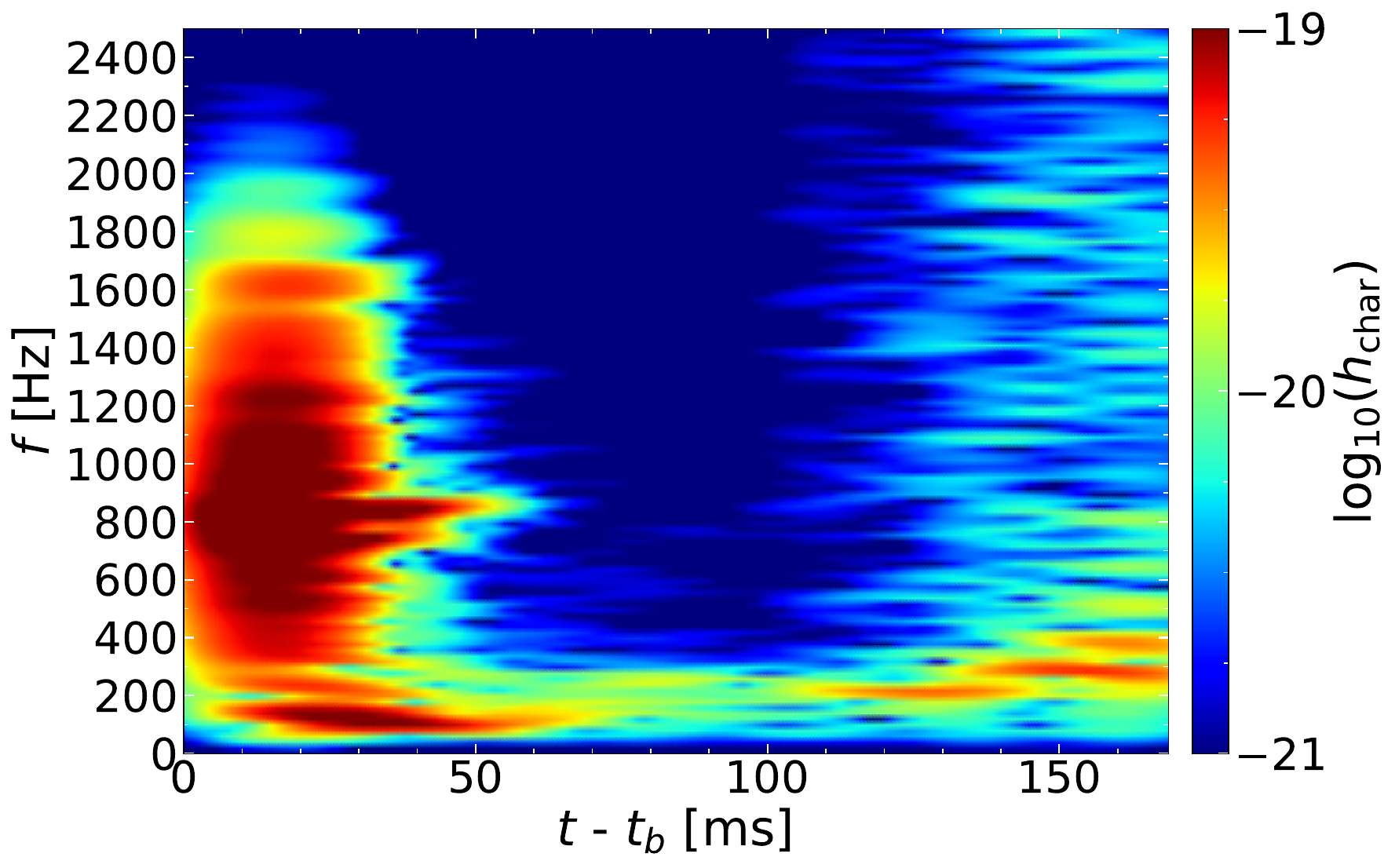}
  \end{subfigure}

  \caption{Spectrograms of the characteristic strains for models $\texttt{25M\_R14B12}$ (first row of
panels) and $\texttt{35M\_R14B12}$ (second row of panels), $\texttt{25M\_R28B12}$ (third row of panels), and $\texttt{35M\_R28B12}$ (fourth row of panels) seen along the pole (left panels) and the equator (right panels) at a source distance of $10\mathrm{kpc}$.}
  \label{fig:spectrogram}
\end{figure*}

Fig.~\ref{fig:spectrogram} shows the spectrograms of the characteristic strain for four models along the polar (left column) and equatorial (right column) directions. All rotating models exhibits a strong $h_\mathrm{char}$ emission across the full frequency range, analogous to the bounce peaks in Fig.~\ref{fig:GW_R14}-\ref{fig:GW_Rest}. The dominant emission occurs near $800\mathrm{Hz}$ in all four cases. The bounce signal is followed by a rising low-frequency component, which gradually increases in strength until the end of the simulation and is confined primarily to $\sim 50 - 400\mathrm{Hz}$. Along the polar direction, we also observe a brief emission across a broad range of frequencies just after bounce, followed by a slowly increasing lower frequency component (between $\sim 50-600 \mathrm{Hz}$) stretching until the end of the simulation. These results are consistent with those of \citet{Shibagaki_GWs_2024}. The spectrograms of the remaining $\texttt{B12}$ models show similar behavior, while the $\texttt{B13}$ models only capture the bounce signal due to their short simulation duration.

\begin{figure}
    \centering
     \begin{subfigure}[b]{0.47\textwidth}
        \centering
        \includegraphics[width = \textwidth]{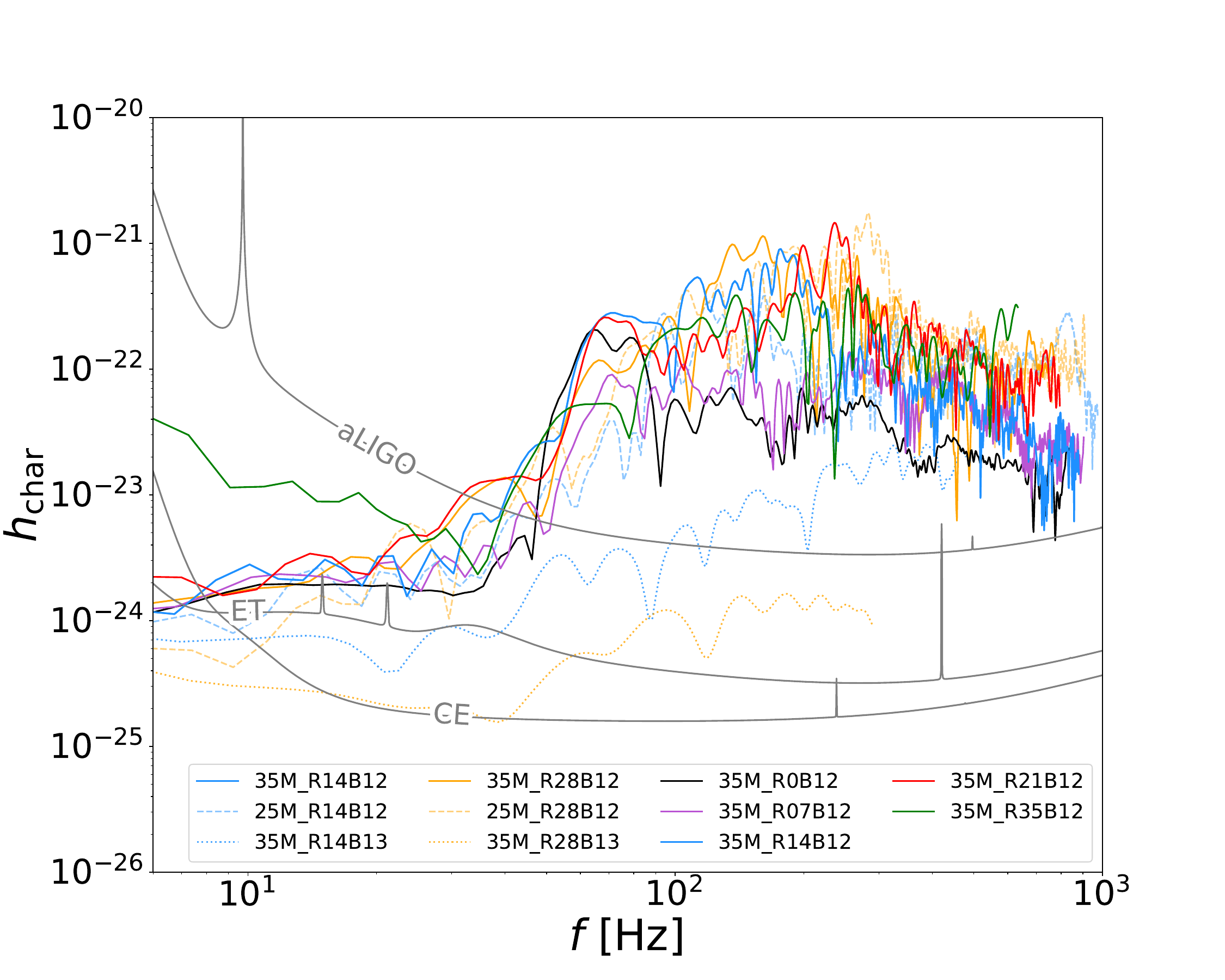}
    \end{subfigure}
     \begin{subfigure}[b]{0.47\textwidth}
        \centering
        \includegraphics[width = \textwidth]{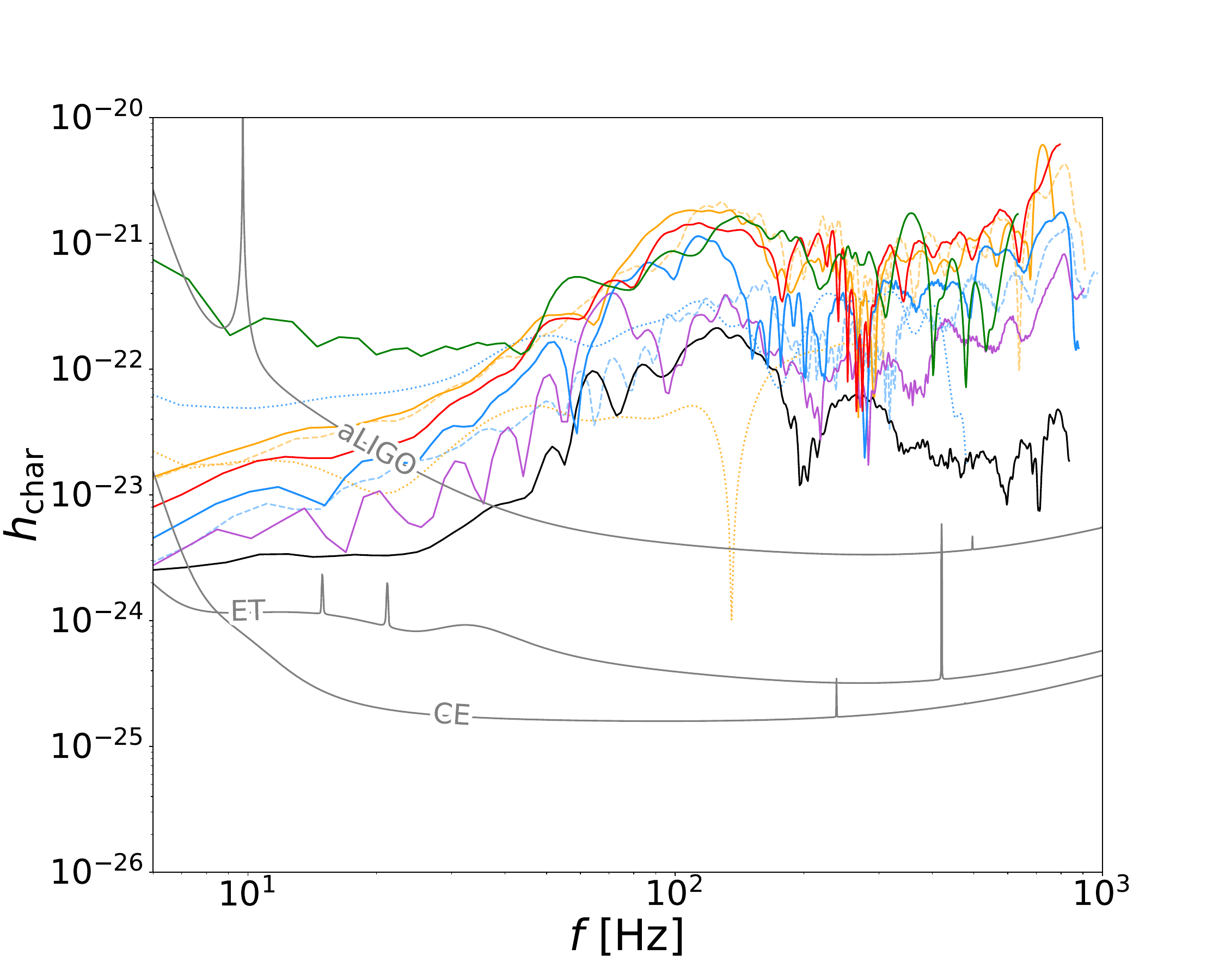}
    \end{subfigure}
    \caption{The top panel shows the polar characteristic strain along as a function of the frequency in Hz for all then models scaled to an observer located at $10\mathrm{kpc}$. The grey lines indicate the predicted sensitivity of aLIGO, Einstein Telescope, and Cosmic Explorer (see \citet{bilby_paper} and \citet{aLIGO}). The bottom panel shows the respective characteristic strain along the equator.}
    \label{fig:ET_ASD}
\end{figure}
In Fig.~\ref{fig:ET_ASD} we show the characteristic GW spectral amplitudes of all models (computed according to Eq.~\ref{eqn:h_char}) at a source distance of $10\,\mathrm{kpc}$ relative to the predicted sensitivity of the Einstein Telescope, aLIGO, and the Cosmic Explorer. Notably, all of our models would be detectable within a certain frequency range, which is in agreement with previous results such as \citet{Ott_2009, Takiwaki_bfield, h_char_Shibagaki2021}, and \citet{Shibagaki_GWs_2024}. The GW signals of these progenitors would in fact even still be visible at a $1-10\mathrm{Mpc}$ distance, as $h_{\mathrm{char}}$ scales with $1/D$.

\section{Discussion and Conclusions}
\label{sec:conclusion}

In this work we investigate the gravitational-wave emission for 10 CCSN progenitor stars. We take 2 progenitors with ZAMS mass $25\M$ and 8 with ZAMS masses of $35\M$ and explore rotation rates between $0.0$ and $3.5 \rs$, and initial magnetic field strengths of $10^{12}\mathrm{G}$ and $10^{13}\mathrm{G}$. We find that the $\texttt{B13}$ models result in a rapidly expanding explosion driven by a highly-collimated jet as the strong poloidal field stabilizes the jet against instabilities~\cite{Moesta_E25}. We overall find that more rapid rotation results in a faster shock expansion, as the faster rotation both amplifies the magnetic fields more efficiently (through rotational winding) and provides a greater amount of rotational energy to power the explosion (see \citet{Shankar2025} for more details on the explosion dynamics). We observe a similar correlation between the rotation rate and the equatorial GW signal in the $+$-polarization: faster rotation yields a more pronounced bounce spike, as the collapsed core is rotationally flattened. Here, the progenitor mass plays a secondary role, resulting in a half as strong bounce peak, but having less impact on the overall signature than the rotation. We find that the highly-collimated $\texttt{B13}$ models result in a high-amplitude $h_+$ signal in the equatorial plane near bounce, while the polar GW signal is suppressed almost completely for both models. Progenitor mass appears to play a more substantial role in determining the PNS evolution, as the $35\M$ models (with the exception of \texttt{35M\_rot28B13}) yield a more massive PNS exceeding that of the $25\M$ by $\sim 0.2-0.3\M$ at the end of the simulations. Overall, our results on the explosion dynamics agree well with the work of \citet{OttAbdikamalov_2012, Shibagaki_GWs_2024}, and \citet{Shankar2025}.

We find that all 10 of our models lie above the detectability threshold for aLIGO, Einstein Telescope, and Cosmic explorer for a source at distance $10\mathrm{kpc}$. With exception of the markedly weaker $\texttt{B13}$ polar $h_{\mathrm{char}}$, these signals would even still be detectable at $1-10\mathrm{Mpc}$, opening the possibility for observing GW emission for CCSNe beyond our galaxy. Since only $\mathcal{O}(1)$ CCSN is expected every 100 years within the Milky Way, it is crucial to assess the prospects of detecting sources outside of this distance. 

While we do not expect $35\M$ progenitor stars at 10\% metallicity in our own galaxy, this choice becomes justified when anticipating the possibility of an extra-galactic CCSNe. Our results indicate that the $35\M$-progenitors would remain detectable for 3rd-generation detectors at source distances of $\sim 1-10\mathrm{Mpc}$, thereby encompassing sources in dwarf galaxies in the local group such as IC 1613, Leo A, WLM, SagDIG, Sextans A, and NGC 3109, where such stars are indeed expected to form (see e.g. \citet{deKoter2014} and \citet{MetalPoorStarsGalaxies_2021}).

Finally, we would like to point out limitations/future work. While the simulations presented here are the largest set of gravitational-wave signatures from 3D GRMHD simulations, we only employ a Leakage/M0 neutrino-transport approximation and only carried out simulations to ~200 ms. This limits the detailed interpretation of the gravitational-wave signature in the later post-bounce evolution. In future work we will carry out new simulations with an M1 neutrino-transport scheme and follow the explosions to later times and investigate the impact of progenitor parameters on additional multimessenger signals such as neutrinos as well as neutrino-induced GWs. We also plan to investigate a wider range of supernova progenitors (e.g. masses and compositions), and their impact on the explosion and associated multimessenger signals. Finally, while most current CCSN simulations still rely on 1D progenitor input, e.g. \citet{Naveen2020}, \citet{Fields2020}, and \citet{Bollig2021} have shown that a 3D progenitor structures can greatly alter the explosion dynamics and morphology. These differences will likely have a substantial impact on the predicted GW signatures.

\section*{Acknowledgements}

PM acknowledges funding through NWO under grant No. OCENW.XL21.XL21.038. This research used resources of the Oak Ridge Leadership Computing Facility at the Oak Ridge National Laboratory, which is supported by the Office of Science of the U.S. Department of Energy under Contract No. DE-AC05-00OR22725. The simulations were carried out on OLCF’s Frontier using allocation AST191. SCS and PM would like to thank Takami Kuroda, Selma de Mink, Stephen Justham, and Matteo Bugli for helpful comments and conversations. 

\section*{Data Availability}

The data underlying this article will be shared on reasonable request to the corresponding author.

\bibliographystyle{mnras}
\bibliography{references}

\begin{thebibliography}{}
\makeatletter
\relax
\def\mn@urlcharsother{\let\do\@makeother \do\$\do\&\do\#\do\^\do\_\do\%\do\~}
\def\mn@doi{\begingroup\mn@urlcharsother \@ifnextchar [ {\mn@doi@} {\mn@doi@[]}}
\def\mn@doi@[#1]#2{\def\@tempa{#1}\ifx\@tempa\@empty \href {http://dx.doi.org/#2} {doi:#2}\else \href {http://dx.doi.org/#2} {#1}\fi \endgroup}
\def\mn@eprint#1#2{\mn@eprint@#1:#2::\@nil}
\def\mn@eprint@arXiv#1{\href {http://arxiv.org/abs/#1} {{\tt arXiv:#1}}}
\def\mn@eprint@dblp#1{\href {http://dblp.uni-trier.de/rec/bibtex/#1.xml} {dblp:#1}}
\def\mn@eprint@#1:#2:#3:#4\@nil{\def\@tempa {#1}\def\@tempb {#2}\def\@tempc {#3}\ifx \@tempc \@empty \let \@tempc \@tempb \let \@tempb \@tempa \fi \ifx \@tempb \@empty \def\@tempb {arXiv}\fi \@ifundefined {mn@eprint@\@tempb}{\@tempb:\@tempc}{\expandafter \expandafter \csname mn@eprint@\@tempb\endcsname \expandafter{\@tempc}}}

\bibitem[\protect\citeauthoryear{Abdikamalov, Pagliaroli  \& Radice}{Abdikamalov et~al.}{2020}]{SNReview}
Abdikamalov E.,  Pagliaroli G.,   Radice D.,  2020, Gravitational Waves from Core-Collapse Supernovae.
Springer Singapore, Singapore, pp 1--37, \mn@doi{10.1007/978-981-15-4702-7_21-1}, \url {https://doi.org/10.1007/978-981-15-4702-7_21-1}

\bibitem[\protect\citeauthoryear{Akiyama, Wheeler, Meier  \& Lichtenstadt}{Akiyama et~al.}{2003a}]{Akiyama_2003}
Akiyama S.,  Wheeler J.~C.,  Meier D.~L.,   Lichtenstadt I.,  2003a, \mn@doi [The Astrophysical Journal] {10.1086/344135}, 584, 954

\bibitem[\protect\citeauthoryear{{Akiyama}, {Wheeler}, {Meier}  \& {Lichtenstadt}}{{Akiyama} et~al.}{2003b}]{MRI}
{Akiyama} S.,  {Wheeler} J.~C.,  {Meier} D.~L.,   {Lichtenstadt} I.,  2003b, \mn@doi [\apj] {10.1086/344135}, \href {https://ui.adsabs.harvard.edu/abs/2003ApJ...584..954A} {584, 954}

\bibitem[\protect\citeauthoryear{Ashton et~al.}{Ashton et~al.}{2019}]{bilby_paper}
Ashton G.,  et~al., 2019, \mn@doi [Astrophys. J. Suppl.] {10.3847/1538-4365/ab06fc}, 241, 27

\bibitem[\protect\citeauthoryear{{Balbus} \& {Hawley}}{{Balbus} \& {Hawley}}{1991}]{Balbus1991}
{Balbus} S.~A.,  {Hawley} J.~F.,  1991, \mn@doi [\apj] {10.1086/170270}, \href {https://ui.adsabs.harvard.edu/abs/1991ApJ...376..214B} {376, 214}

\bibitem[\protect\citeauthoryear{Barsotti, Fritschel, Evans  \& Gras}{Barsotti et~al.}{2018}]{aLIGO}
Barsotti L.,  Fritschel P.,  Evans M.,   Gras S.,  2018, Technical Report LIGO-T1800044-v5, Updated Advanced LIGO Sensitivity Design Curve, \url {https://dcc.ligo.org/LIGO-T1800044/public}.
LIGO Laboratory, \url {https://dcc.ligo.org/LIGO-T1800044/public}

\bibitem[\protect\citeauthoryear{{Bethe} \& {Wilson}}{{Bethe} \& {Wilson}}{1985}]{BetheWilson1985}
{Bethe} H.~A.,  {Wilson} J.~R.,  1985, \mn@doi [\apj] {10.1086/163343}, \href {https://ui.adsabs.harvard.edu/abs/1985ApJ...295...14B} {295, 14}

\bibitem[\protect\citeauthoryear{{Bisnovatyi-Kogan}}{{Bisnovatyi-Kogan}}{1971}]{Bisnovatyi_Kogan_1971}
{Bisnovatyi-Kogan} G.~S.,  1971, \sovast, \href {https://ui.adsabs.harvard.edu/abs/1971SvA....14..652B} {14, 652}

\bibitem[\protect\citeauthoryear{{Bollig}, {Yadav}, {Kresse}, {Janka}, {M{\"u}ller}  \& {Heger}}{{Bollig} et~al.}{2021}]{Bollig2021}
{Bollig} R.,  {Yadav} N.,  {Kresse} D.,  {Janka} H.-T.,  {M{\"u}ller} B.,   {Heger} A.,  2021, \mn@doi [\apj] {10.3847/1538-4357/abf82e}, \href {https://ui.adsabs.harvard.edu/abs/2021ApJ...915...28B} {915, 28}

\bibitem[\protect\citeauthoryear{{Bugli}, {Guilet}, {Foglizzo}  \& {Obergaulinger}}{{Bugli} et~al.}{2023}]{Bugli2023}
{Bugli} M.,  {Guilet} J.,  {Foglizzo} T.,   {Obergaulinger} M.,  2023, \mn@doi [\mnras] {10.1093/mnras/stad496}, \href {https://ui.adsabs.harvard.edu/abs/2023MNRAS.520.5622B} {520, 5622}

\bibitem[\protect\citeauthoryear{Burrows \& Vartanyan}{Burrows \& Vartanyan}{2021}]{Burrows_2021}
Burrows A.,  Vartanyan D.,  2021, \mn@doi [Nature] {10.1038/s41586-020-03059-w}, 589, 29–39

\bibitem[\protect\citeauthoryear{{Burrows}, {Dessart}, {Livne}, {Ott}  \& {Murphy}}{{Burrows} et~al.}{2007}]{Burrows2007}
{Burrows} A.,  {Dessart} L.,  {Livne} E.,  {Ott} C.~D.,   {Murphy} J.,  2007, \mn@doi [\apj] {10.1086/519161}, \href {https://ui.adsabs.harvard.edu/abs/2007ApJ...664..416B} {664, 416}

\bibitem[\protect\citeauthoryear{Einfeldt}{Einfeldt}{1988}]{HLLE}
Einfeldt B.,  1988, in Proceedings of the 16th International Symposium on Shock Tubes and Waves. VCH Verlag, Aachen, Germany, p.~671

\bibitem[\protect\citeauthoryear{{Fields} \& {Couch}}{{Fields} \& {Couch}}{2020}]{Fields2020}
{Fields} C.~E.,  {Couch} S.~M.,  2020, \mn@doi [\apj] {10.3847/1538-4357/abada7}, \href {https://ui.adsabs.harvard.edu/abs/2020ApJ...901...33F} {901, 33}

\bibitem[\protect\citeauthoryear{{Garcia} et~al.,}{{Garcia} et~al.}{2021}]{MetalPoorStarsGalaxies_2021}
{Garcia} M.,  et~al., 2021, \mn@doi [Experimental Astronomy] {10.1007/s10686-021-09785-x}, \href {https://ui.adsabs.harvard.edu/abs/2021ExA....51..887G} {51, 887}

\bibitem[\protect\citeauthoryear{{Heger}, {Langer}  \& {Woosley}}{{Heger} et~al.}{2000}]{E25}
{Heger} A.,  {Langer} N.,   {Woosley} S.~E.,  2000, \mn@doi [\apj] {10.1086/308158}, \href {https://ui.adsabs.harvard.edu/abs/2000ApJ...528..368H} {528, 368}

\bibitem[\protect\citeauthoryear{{Janka}}{{Janka}}{2012}]{Janka_Neutrino_Mechanism}
{Janka} H.-T.,  2012, \mn@doi [Annual Review of Nuclear and Particle Science] {10.1146/annurev-nucl-102711-094901}, \href {https://ui.adsabs.harvard.edu/abs/2012ARNPS..62..407J} {62, 407}

\bibitem[\protect\citeauthoryear{Janka, Langanke, Marek, Martinezpindeo  \& Müller}{Janka et~al.}{2007}]{Janka_2007}
Janka H.~T.,  Langanke K.,  Marek A.,  Martinezpindeo G.,   Müller B.,  2007, \mn@doi [Physics Reports] {10.1016/j.physrep.2007.02.002}, 442, 38–74

\bibitem[\protect\citeauthoryear{{Kuroda}, {Takiwaki}  \& {Kotake}}{{Kuroda} et~al.}{2014}]{KurodaTaKo_14}
{Kuroda} T.,  {Takiwaki} T.,   {Kotake} K.,  2014, \mn@doi [\prd] {10.1103/PhysRevD.89.044011}, \href {https://ui.adsabs.harvard.edu/abs/2014PhRvD..89d4011K} {89, 044011}

\bibitem[\protect\citeauthoryear{{Lattimer} \& {Swesty}}{{Lattimer} \& {Swesty}}{1991}]{LS220}
{Lattimer} J.~M.,  {Swesty} D.~F.,  1991, \mn@doi [\nphysa] {10.1016/0375-9474(91)90452-C}, \href {https://ui.adsabs.harvard.edu/abs/1991NuPhA.535..331L} {535, 331}

\bibitem[\protect\citeauthoryear{{LeBlanc} \& {Wilson}}{{LeBlanc} \& {Wilson}}{1970}]{LeBlanc_Wilson_1970}
{LeBlanc} J.~M.,  {Wilson} J.~R.,  1970, \mn@doi [\apj] {10.1086/150558}, \href {https://ui.adsabs.harvard.edu/abs/1970ApJ...161..541L} {161, 541}

\bibitem[\protect\citeauthoryear{{Masada}, {Takiwaki}  \& {Kotake}}{{Masada} et~al.}{2022}]{Masada_2022}
{Masada} Y.,  {Takiwaki} T.,   {Kotake} K.,  2022, \mn@doi [\apj] {10.3847/1538-4357/ac34f6}, \href {https://ui.adsabs.harvard.edu/abs/2022ApJ...924...75M} {924, 75}

\bibitem[\protect\citeauthoryear{{Meier}, {Epstein}, {Arnett}  \& {Schramm}}{{Meier} et~al.}{1976}]{Meier1976}
{Meier} D.~L.,  {Epstein} R.~I.,  {Arnett} W.~D.,   {Schramm} D.~N.,  1976, \mn@doi [\apj] {10.1086/154235}, \href {https://ui.adsabs.harvard.edu/abs/1976ApJ...204..869M} {204, 869}

\bibitem[\protect\citeauthoryear{{Moenchmeyer}, {Schaefer}, {Mueller}  \& {Kates}}{{Moenchmeyer} et~al.}{1991}]{RotLaw_1991}
{Moenchmeyer} R.,  {Schaefer} G.,  {Mueller} E.,   {Kates} R.~E.,  1991, \aap, \href {https://ui.adsabs.harvard.edu/abs/1991A&A...246..417M} {246, 417}

\bibitem[\protect\citeauthoryear{{M{\"o}sta} et~al.,}{{M{\"o}sta} et~al.}{2014a}]{Kink_instability_2014}
{M{\"o}sta} P.,  et~al., 2014a, \mn@doi [\apjl] {10.1088/2041-8205/785/2/L29}, \href {https://ui.adsabs.harvard.edu/abs/2014ApJ...785L..29M} {785, L29}

\bibitem[\protect\citeauthoryear{{M{\"o}sta} et~al.,}{{M{\"o}sta} et~al.}{2014b}]{Moesta_E25}
{M{\"o}sta} P.,  et~al., 2014b, \mn@doi [\apjl] {10.1088/2041-8205/785/2/L29}, \href {https://ui.adsabs.harvard.edu/abs/2014ApJ...785L..29M} {785, L29}

\bibitem[\protect\citeauthoryear{M{\"o}sta, Ott, Radice, Roberts, Schnetter  \& Haas}{M{\"o}sta et~al.}{2015}]{Moesta_2015_Dynamo}
M{\"o}sta P.,  Ott C.~D.,  Radice D.,  Roberts L.~F.,  Schnetter E.,   Haas R.,  2015, \mn@doi [Nature] {10.1038/nature15755}, 528, 376

\bibitem[\protect\citeauthoryear{{M{\"o}sta}, {Roberts}, {Halevi}, {Ott}, {Lippuner}, {Haas}  \& {Schnetter}}{{M{\"o}sta} et~al.}{2018}]{Mösta2018}
{M{\"o}sta} P.,  {Roberts} L.~F.,  {Halevi} G.,  {Ott} C.~D.,  {Lippuner} J.,  {Haas} R.,   {Schnetter} E.,  2018, \mn@doi [\apj] {10.3847/1538-4357/aad6ec}, \href {https://ui.adsabs.harvard.edu/abs/2018ApJ...864..171M} {864, 171}

\bibitem[\protect\citeauthoryear{{M{\"u}ller}, {Janka}  \& {Wongwathanarat}}{{M{\"u}ller} et~al.}{2012}]{NeutrinoDriven2_Müller}
{M{\"u}ller} E.,  {Janka} H.~T.,   {Wongwathanarat} A.,  2012, \mn@doi [\aap] {10.1051/0004-6361/201117611}, \href {https://ui.adsabs.harvard.edu/abs/2012A&A...537A..63M} {537, A63}

\bibitem[\protect\citeauthoryear{{O'Connor} \& {Ott}}{{O'Connor} \& {Ott}}{2011}]{OConnor2011}
{O'Connor} E.,  {Ott} C.~D.,  2011, \mn@doi [\apj] {10.1088/0004-637X/730/2/70}, \href {https://ui.adsabs.harvard.edu/abs/2011ApJ...730...70O} {730, 70}

\bibitem[\protect\citeauthoryear{Obergaulinger \& Aloy}{Obergaulinger \& Aloy}{2021}]{ObergaulingerAloy2021}
Obergaulinger M.,  Aloy M.~A.,  2021, \mn@doi [Monthly Notices of the Royal Astronomical Society] {10.1093/mnras/stab295}, 503, 4942

\bibitem[\protect\citeauthoryear{{Obergaulinger}, {Cerd{\'a}-Dur{\'a}n}, {M{\"u}ller}  \& {Aloy}}{{Obergaulinger} et~al.}{2009}]{Obergaulinger_2009}
{Obergaulinger} M.,  {Cerd{\'a}-Dur{\'a}n} P.,  {M{\"u}ller} E.,   {Aloy} M.~A.,  2009, \mn@doi [\aap] {10.1051/0004-6361/200811323}, \href {https://ui.adsabs.harvard.edu/abs/2009A&A...498..241O} {498, 241}

\bibitem[\protect\citeauthoryear{Ott}{Ott}{2009}]{Ott_2009}
Ott C.~D.,  2009, \mn@doi [Classical and Quantum Gravity] {10.1088/0264-9381/26/6/063001}, 26, 063001

\bibitem[\protect\citeauthoryear{{Ott} et~al.,}{{Ott} et~al.}{2012}]{OttAbdikamalov_2012}
{Ott} C.~D.,  et~al., 2012, \mn@doi [\prd] {10.1103/PhysRevD.86.024026}, \href {https://ui.adsabs.harvard.edu/abs/2012PhRvD..86b4026O} {86, 024026}

\bibitem[\protect\citeauthoryear{{Packet}}{{Packet}}{1981}]{Packet1981_RapidRotation}
{Packet} W.,  1981, \aap, \href {https://ui.adsabs.harvard.edu/abs/1981A&A...102...17P} {102, 17}

\bibitem[\protect\citeauthoryear{Radice, Galeazzi, Lippuner, Roberts, Ott  \& Rezzolla}{Radice et~al.}{2016}]{Radice_2016}
Radice D.,  Galeazzi F.,  Lippuner J.,  Roberts L.~F.,  Ott C.~D.,   Rezzolla L.,  2016, \mn@doi [Monthly Notices of the Royal Astronomical Society] {10.1093/mnras/stw1227}, 460, 3255

\bibitem[\protect\citeauthoryear{Radice, Perego, Hotokezaka, Fromm, Bernuzzi  \& Roberts}{Radice et~al.}{2018}]{Radice_2018}
Radice D.,  Perego A.,  Hotokezaka K.,  Fromm S.~A.,  Bernuzzi S.,   Roberts L.~F.,  2018, \mn@doi [The Astrophysical Journal] {10.3847/1538-4357/aaf054}, 869, 130

\bibitem[\protect\citeauthoryear{Raynaud, Guilet, Janka  \& Gastine}{Raynaud et~al.}{2020}]{Raynaud2020}
Raynaud R.,  Guilet J.,  Janka H.-T.,   Gastine T.,  2020, \mn@doi [Science Advances] {10.1126/sciadv.aay2732}, 6, eaay2732

\bibitem[\protect\citeauthoryear{{Reboul-Salze}, {Guilet}, {Raynaud}  \& {Bugli}}{{Reboul-Salze} et~al.}{2021}]{Reboul-Salze_2021}
{Reboul-Salze} A.,  {Guilet} J.,  {Raynaud} R.,   {Bugli} M.,  2021, \mn@doi [\aap] {10.1051/0004-6361/202038369}, \href {https://ui.adsabs.harvard.edu/abs/2021A&A...645A.109R} {645, A109}

\bibitem[\protect\citeauthoryear{{Reisswig}, {Ott}, {Sperhake}  \& {Schnetter}}{{Reisswig} et~al.}{2011}]{QuadrupoleExtractionMethod}
{Reisswig} C.,  {Ott} C.~D.,  {Sperhake} U.,   {Schnetter} E.,  2011, \mn@doi [\prd] {10.1103/PhysRevD.83.064008}, \href {https://ui.adsabs.harvard.edu/abs/2011PhRvD..83f4008R} {83, 064008}

\bibitem[\protect\citeauthoryear{{Reisswig}, {Haas}, {Ott}, {Abdikamalov}, {M{\"o}sta}, {Pollney}  \& {Schnetter}}{{Reisswig} et~al.}{2013}]{WENO5_1}
{Reisswig} C.,  {Haas} R.,  {Ott} C.~D.,  {Abdikamalov} E.,  {M{\"o}sta} P.,  {Pollney} D.,   {Schnetter} E.,  2013, \mn@doi [\prd] {10.1103/PhysRevD.87.064023}, \href {https://ui.adsabs.harvard.edu/abs/2013PhRvD..87f4023R} {87, 064023}

\bibitem[\protect\citeauthoryear{{Scheidegger}, {Whitehouse}, {K{\"a}ppeli}  \& {Liebend{\"o}rfer}}{{Scheidegger} et~al.}{2010a}]{Scheidegger_E_GW}
{Scheidegger} S.,  {Whitehouse} S.~C.,  {K{\"a}ppeli} R.,   {Liebend{\"o}rfer} M.,  2010a, \mn@doi [Classical and Quantum Gravity] {10.1088/0264-9381/27/11/114101}, \href {https://ui.adsabs.harvard.edu/abs/2010CQGra..27k4101S} {27, 114101}

\bibitem[\protect\citeauthoryear{{Scheidegger}, {K{\"a}ppeli}, {Whitehouse}, {Fischer}  \& {Liebend{\"o}rfer}}{{Scheidegger} et~al.}{2010b}]{ScheideggerKWF_10}
{Scheidegger} S.,  {K{\"a}ppeli} R.,  {Whitehouse} S.~C.,  {Fischer} T.,   {Liebend{\"o}rfer} M.,  2010b, \mn@doi [\aap] {10.1051/0004-6361/200913220}, \href {https://ui.adsabs.harvard.edu/abs/2010A&A...514A..51S} {514, A51}

\bibitem[\protect\citeauthoryear{{Schneider}, {Ohlmann}, {Podsiadlowski}, {R{\"o}pke}, {Balbus}, {Pakmor}  \& {Springel}}{{Schneider} et~al.}{2024}]{MagneticFieldsMassiveStars}
{Schneider} F. R.~N.,  {Ohlmann} S.~T.,  {Podsiadlowski} P.,  {R{\"o}pke} F.~K.,  {Balbus} S.~A.,  {Pakmor} R.,   {Springel} V.,  2024, in {Mackey} J.,  {Vink} J.~S.,   {St-Louis} N.,  eds,  IAU Symposium Vol. 361, IAU Symposium. pp 212--217, \mn@doi{10.1017/S1743921322002794}

\bibitem[\protect\citeauthoryear{{Shankar}, {M{\"o}sta}, {Brandt}, {Haas}, {Schnetter}  \& {de Graaf}}{{Shankar} et~al.}{2023}]{Gram-X}
{Shankar} S.,  {M{\"o}sta} P.,  {Brandt} S.~R.,  {Haas} R.,  {Schnetter} E.,   {de Graaf} Y.,  2023, \mn@doi [Classical and Quantum Gravity] {10.1088/1361-6382/acf2d9}, \href {https://ui.adsabs.harvard.edu/abs/2023CQGra..40t5009S} {40, 205009}

\bibitem[\protect\citeauthoryear{{Shankar}, {M{\"o}sta}, {Haas}  \& {Schnetter}}{{Shankar} et~al.}{2025}]{Shankar2025}
{Shankar} S.,  {M{\"o}sta} P.,  {Haas} R.,   {Schnetter} E.,  2025, \mn@doi [arXiv e-prints] {10.48550/arXiv.2504.11537}, \href {https://ui.adsabs.harvard.edu/abs/2025arXiv250411537S} {p. arXiv:2504.11537}

\bibitem[\protect\citeauthoryear{{Shibagaki}, {Kuroda}, {Kotake}  \& {Takiwaki}}{{Shibagaki} et~al.}{2021a}]{Kuroda_2021}
{Shibagaki} S.,  {Kuroda} T.,  {Kotake} K.,   {Takiwaki} T.,  2021a, \mn@doi [\mnras] {10.1093/mnras/stab228}, \href {https://ui.adsabs.harvard.edu/abs/2021MNRAS.502.3066S} {502, 3066}

\bibitem[\protect\citeauthoryear{Shibagaki, Kuroda, Kotake  \& Takiwaki}{Shibagaki et~al.}{2021b}]{h_char_Shibagaki2021}
Shibagaki S.,  Kuroda T.,  Kotake K.,   Takiwaki T.,  2021b, \mn@doi [Monthly Notices of the Royal Astronomical Society] {10.1093/mnras/stab228}, 502, 3066

\bibitem[\protect\citeauthoryear{{Shibagaki}, {Kuroda}, {Kotake}, {Takiwaki}  \& {Fischer}}{{Shibagaki} et~al.}{2024}]{Shibagaki_GWs_2024}
{Shibagaki} S.,  {Kuroda} T.,  {Kotake} K.,  {Takiwaki} T.,   {Fischer} T.,  2024, \mn@doi [\mnras] {10.1093/mnras/stae1361}, \href {https://ui.adsabs.harvard.edu/abs/2024MNRAS.531.3732S} {531, 3732}

\bibitem[\protect\citeauthoryear{Shibata, Liu, Shapiro  \& Stephens}{Shibata et~al.}{2006}]{Shibata2006}
Shibata M.,  Liu Y.~T.,  Shapiro S.~L.,   Stephens B.~C.,  2006, \mn@doi [Phys. Rev. D] {10.1103/PhysRevD.74.104026}, 74, 104026

\bibitem[\protect\citeauthoryear{{Takiwaki} \& {Kotake}}{{Takiwaki} \& {Kotake}}{2011}]{Takiwaki_bfield}
{Takiwaki} T.,  {Kotake} K.,  2011, \mn@doi [\apj] {10.1088/0004-637X/743/1/30}, \href {https://ui.adsabs.harvard.edu/abs/2011ApJ...743...30T} {743, 30}

\bibitem[\protect\citeauthoryear{{Takiwaki}, {Kotake}, {Nagataki}  \& {Sato}}{{Takiwaki} et~al.}{2004}]{Takiwaki_rotation_law}
{Takiwaki} T.,  {Kotake} K.,  {Nagataki} S.,   {Sato} K.,  2004, \mn@doi [\apj] {10.1086/424993}, \href {https://ui.adsabs.harvard.edu/abs/2004ApJ...616.1086T} {616, 1086}

\bibitem[\protect\citeauthoryear{Tchekhovskoy, McKinney  \& Narayan}{Tchekhovskoy et~al.}{2007}]{WENO5_2}
Tchekhovskoy A.,  McKinney J.~C.,   Narayan R.,  2007, \mn@doi [Monthly Notices of the Royal Astronomical Society] {10.1111/j.1365-2966.2007.11876.x}, 379, 469

\bibitem[\protect\citeauthoryear{{Thompson} \& {Duncan}}{{Thompson} \& {Duncan}}{1993}]{Thompson_1993}
{Thompson} C.,  {Duncan} R.~C.,  1993, \mn@doi [\apj] {10.1086/172580}, \href {https://ui.adsabs.harvard.edu/abs/1993ApJ...408..194T} {408, 194}

\bibitem[\protect\citeauthoryear{Toro}{Toro}{1999}]{Toro1999}
Toro E.~F.,  1999, Riemann Solvers and Numerical Methods for Fluid Dynamics.
Springer, Berlin

\bibitem[\protect\citeauthoryear{{T{\'o}th}}{{T{\'o}th}}{2000}]{Toth}
{T{\'o}th} G.,  2000, \mn@doi [Journal of Computational Physics] {10.1006/jcph.2000.6519}, \href {https://ui.adsabs.harvard.edu/abs/2000JCoPh.161..605T} {161, 605}

\bibitem[\protect\citeauthoryear{{Tramper}, {Sana}, {de Koter}, {Kaper}  \& {Ram{\'\i}rez-Agudelo}}{{Tramper} et~al.}{2014}]{deKoter2014}
{Tramper} F.,  {Sana} H.,  {de Koter} A.,  {Kaper} L.,   {Ram{\'\i}rez-Agudelo} O.~H.,  2014, \mn@doi [\aap] {10.1051/0004-6361/201424312}, \href {https://ui.adsabs.harvard.edu/abs/2014A&A...572A..36T} {572, A36}

\bibitem[\protect\citeauthoryear{{Vartanyan}, {Burrows}, {Radice}, {Skinner}  \& {Dolence}}{{Vartanyan} et~al.}{2019}]{NeutrinoDriven_Vartanyan}
{Vartanyan} D.,  {Burrows} A.,  {Radice} D.,  {Skinner} M.~A.,   {Dolence} J.,  2019, \mn@doi [\mnras] {10.1093/mnras/sty2585}, \href {https://ui.adsabs.harvard.edu/abs/2019MNRAS.482..351V} {482, 351}

\bibitem[\protect\citeauthoryear{Wheeler, Meier  \& Wilson}{Wheeler et~al.}{2002}]{Wheeler_2002}
Wheeler J.~C.,  Meier D.~L.,   Wilson J.~R.,  2002, \mn@doi [The Astrophysical Journal] {10.1086/338953}, 568, 807

\bibitem[\protect\citeauthoryear{{Woosley} \& {Heger}}{{Woosley} \& {Heger}}{2006a}]{WoosleyHeger2006}
{Woosley} S.~E.,  {Heger} A.,  2006a, \mn@doi [\apj] {10.1086/498500}, \href {https://ui.adsabs.harvard.edu/abs/2006ApJ...637..914W} {637, 914}

\bibitem[\protect\citeauthoryear{{Woosley} \& {Heger}}{{Woosley} \& {Heger}}{2006b}]{35OC}
{Woosley} S.~E.,  {Heger} A.,  2006b, \mn@doi [\apj] {10.1086/498500}, \href {https://ui.adsabs.harvard.edu/abs/2006ApJ...637..914W} {637, 914}

\bibitem[\protect\citeauthoryear{{Yadav}, {M{\"u}ller}, {Janka}, {Melson}  \& {Heger}}{{Yadav} et~al.}{2020}]{Naveen2020}
{Yadav} N.,  {M{\"u}ller} B.,  {Janka} H.~T.,  {Melson} T.,   {Heger} A.,  2020, \mn@doi [\apj] {10.3847/1538-4357/ab66bb}, \href {https://ui.adsabs.harvard.edu/abs/2020ApJ...890...94Y} {890, 94}

\bibitem[\protect\citeauthoryear{{de Blank} \& {Schep}}{{de Blank} \& {Schep}}{1991}]{Kink_instab_1991}
{de Blank} H.~J.,  {Schep} T.~J.,  1991, \mn@doi [Physics of Fluids B] {10.1063/1.859805}, \href {https://ui.adsabs.harvard.edu/abs/1991PhFlB...3.1136D} {3, 1136}

\bibitem[\protect\citeauthoryear{{de Mink}, {Cantiello}, {Langer}  \& {Pols}}{{de Mink} et~al.}{2010}]{Chemically_Homogeneous_Evolution_deMink_2010}
{de Mink} S.~E.,  {Cantiello} M.,  {Langer} N.,   {Pols} O.~R.,  2010, in {Kalogera} V.,  {van der Sluys} M.,  eds,  American Institute of Physics Conference Series Vol. 1314, International Conference on Binaries: in celebration of Ron Webbink's 65th Birthday. AIP, pp 291--296 (\mn@eprint {arXiv} {1010.2177}), \mn@doi{10.1063/1.3536387}

\makeatother
\end{thebibliography}

\bsp	
\label{lastpage}
\end{document}